\documentclass[12pt,a4paper]{article}
\usepackage{epsfig,float}
\begin{document}
\renewcommand{\textfraction}{0.1}
\renewcommand{\thefootnote}{\fnsymbol{footnote}}
\title{ Nucleon-Nucleon Optical Model for Energies to 3 GeV}
\author{ A. Funk$^1$\footnote{funk@physnet2.uni-hamburg.de}, H.V. von
  Geramb$^{1,2}$
\footnote{geramb@uni-hamburg.de} and  K.A. Amos$^{2}$
\footnote{amos@higgs.ph.unimelb.edu.au}\\[0.3cm] $^1$ 
Theoretische Kernphysik, Universit\"at Hamburg,\\
Luruper Chaussee 149, D-22761 Hamburg, Germany\\[0.3cm] $^2$ School of
Physics, University of Melbourne,\\
Victoria 3010,  Australia}
\sloppy

\newcommand{\fdag}{\not\kern-.25em}
\maketitle
\begin{abstract}
Several nucleon-nucleon potentials,  Paris, Nijmegen, Argonne, 
and those derived by quantum inversion,
which  describe the  NN interaction for  $T_{Lab}\le 300$ 
MeV are extended in their  range of application 
as  NN optical models. Extensions are made in r-space using
complex  separable  potentials definable with a wide range 
of form factor options including those of
boundary condition models. We use the latest phase shift  
analyses SP00 (FA00, WI00) of Arndt {\em et al.} from 300 MeV to 3 GeV
to determine these extensions.  The imaginary parts of the 
optical model interactions
account for loss of flux into direct or resonant production processes.
The optical potential approach is of particular value as it
 permits one to visualize fusion, 
and subsequent fission, of nucleons when $T_{Lab}>2$ GeV. We do so by
 calculating the 
scattering  wave functions to specify the energy and radial 
dependences of flux losses  and of probability distributions. Furthermore, 
half-off the energy shell t-matrices are presented as they are readily 
deduced with this approach.
Such t-matrices are required for studies  of few- and many-body nuclear 
reactions.
\end{abstract}


\section{Introduction}

A theoretical description of nucleon-nucleon (NN)  scattering  is
a fundamental ingredient  for the understanding of nuclear structure
and scattering  of few- and many-body nuclear systems 
\cite{Fes92,Mac89,Mac01a}. 
This is a paradigm of nuclear physics. Of the spectrum, 
low energy NN scattering
traditionally is described in terms of few degrees of
freedom of which  spin and isospin symmetries play the predominant role. 
At medium energies, production processes and inelasticities
become important and several elementary systems composed of nucleons 
and mesons contribute to NN scattering. While these nucleons and
mesons are emergent structures from QCD, at present there is no
quantitative description of NN  scattering above the inelastic threshold
either in terms of QCD or of the emergent 
nucleons and mesons \cite{Myh88}. 

Theoretically undisputed is the need for relativity \cite{Pol98} 
of which there are two aspects. 
First is the increasing importance of relativistic kinematics
as the kinetic energy becomes comparable to the rest masses of the
scattering particles. Second, particle production is inherently 
relativistic requiring, ultimately, a  description in terms of 
highly non-linear QCD. But that non-linearity inhibits a facile 
QCD explanation of NN  scattering. Notwithstanding, there exist 
hybrid models which offset that non-linearity
in  seeking explanation of the excitation spectra and of the 
scattering  of hadrons \cite{Kuk98,Jen00}. All use heavy valence  
quarks, with an effective mass typically  of 300 MeV, and massive Goldstone
bosons in lieu of massless gluons. They also maintain color degrees of
freedom. As well there are  effective quantum field theories (EQF)
that link the quark-gluon structure of the standard model to low
energy nuclear physics \cite{Sca97,Ric00}. 
Currently these latter approaches are very popular
as they may give a foundation and interpretation of emergent structures.
But like the hybrid models, due to the underlying expansion
schemes used with EQF, many {\em ad hoc} degrees of freedom are involved.

The experimental NN data and its parameterization
in terms of amplitudes and phase shifts, are very smooth with energy 
to 3 GeV
\cite{Arn83,Bys87,GWU00}; a feature
which supports use of the {\em classic} approach using  
a free NN interaction potential. By so doing one uses a minimal number of
degrees of freedom with  again those degrees of freedom being associated
with the spin and isospin of the total system. Of course, 
this {\it classic} approach sacrifices all reliance on substructures. 
However, 
the underlying dynamic still reflects  its geometric
facet by means of surfaces and boundary conditions.
The success of bag models is direct
evidence of the crucial role such boundary conditions play with the
emergent structures from them being  direct consequences of
QCD confinement. This is further
support for our view that an explanation of elastic NN  
scattering need not,
if will not,  depend explicitly upon QCD details. Only 
geometric attributes 
of the QCD domain, we contend, are required to explain 
most data.
This view is well supported by high energy scattering for 
which the geometric limits
of the S-matrix are reached and form factors are defined 
independent of energy.
In the transition region the geometric limits are not
reached and the factorization schemes \cite{Lo87,Mat94} 
used at 
higher energies do not apply.
   
Of course, in the last decade or so, there have been 
several theoretical attempts  built upon boson exchange 
models to explain NN  
scattering data to 1 GeV. All such have given but 
qualitative results, often requiring many degrees of 
freedom even to achieve 
that qualitative agreement and  despite explicit inclusion of 
$\Delta$ and $N^*$  resonances. 
Optical model studies have also been made for medium and modest  
high energy NN 
scattering \cite{Neu91,Ger98,Ger00}, and they can  be improved 
to give a high
quality  description of scattering at medium energy. 

A  high quality fit of on-shell t-matrices by a potential model is very
desirable  also as it facilitates  extension into the off-shell domain; 
properties which are needed in  few and many body calculations.
In particular, microscopic optical
model potentials for elastic nucleon-nucleus scattering and
bremsstrahlung reactions that give quantitative results, 
require a careful and exact treatment of the off-shell NN t-matrices 
\cite{Amo00}. 
Furthermore,  calculations of such entities have shown that 
it is crucial to have 
on-shell values of the t-matrices in best possible agreement with 
NN  data at all energies. Concomitantly one needs high 
precision NN  data against which one can specify NN interactions.

There are many studies of few and many body problems
in the low energy regime  $T_{Lab} <300$ MeV and the results 
have consequences for any model extension above threshold \cite{Mac01a}. 
We note in this context that significant off-shell
differences in  t-matrices are known to exist between
the theoretically well motivated boson exchange models of NN 
scattering in this regime. It remains difficult to attribute with certainty
any particular dynamical or kinematical feature with those differences.
Non-locality, explicit energy dependence and features  associated
with relativistic kinematics are some possibilities.
 
In contrast, there is the quantum inverse scattering approach by 
which any on-shell t-matrix can be continued into the off-shell domain 
\cite{Cha77}. A specific method is the Gel'fand--Levitan--Marchenko 
inversion algorithm for Sturm--Liouville equations. 
This approach to specify t-matrices off-shell is appropriate when the 
physical S-matrix is unitary and the equation of motion is of the
Sturm--Liouville type. Such  is valid without modification  for 
NN  t-matrices in the
energy regime to 300 MeV. Mathematically, the Gel'fand--Levitan--Marchenko
algorithm is a method based upon a class of real and regular potentials.
In the spirit of inverse scattering, we generalize that method
for non-unitary S-matrices. By that means we  generate an
NN  optical model separately  for each partial wave. 
The  algorithm we  have developed allows studies of
complex separable potentials in combination with any  background  
potential. The background potential can be  any of the 
existing r-space {\em NN} potentials. We have not used
k-space background potentials, such as Bonn-B \cite{Mac89}, 
Bonn-CD \cite{Mac01} 
and OSBEP \cite{Jae98}, albeit that  similar analyses can be  
made with them.

We limit the reference potential  to the well known real r-space 
potentials from Paris \cite{Lac80}, Nijmegen \cite{Sto93} 
(Reid93, Nijmegen-I, Nijmegen-II), Argonne \cite{Wir95} (AV18),
and from inversion \cite{Kir89,San97}. To them we add channel 
dependent complex
separable potentials with  energy dependent strengths. 
For given input data results then,   the full potentials 
are unique. The experimental background and motivation for  
analyses using an
optical model is given in Sect.\,II. A detailed description of   
the theoretical  algorithm is given in Sect.\,III.
A discussion of results is  given
in Sect.\,IV while a  summary is given in Sect.\,V.
                                                             
\section{Survey of data and motivation for the optical potential}

NN scattering is a long standing problem
which  has been reviewed often as the database developed 
\cite{Arn83,Bys87,Mac01,PWA93}. 
The low energy data has been analyzed by the VPI/GWU group
\cite{GWU00} for $T_{Lab}\leq400$ MeV, the  Nijmegen group \cite{PWA93} 
with the  NN phase shift results  PWA93 for 
$T_{Lab}\leq 350$ MeV, and by Machleidt \cite{Mac01} giving the  
Bonn-CD-2000. 
Of these, the VPI/GWU group has given many solutions for this low 
energy regime over the years, all of which  have been listed by 
Arndt {\em et al.} in a very recent publication \cite{GWU00}.
For their use  note that the solution name reflects the season 
and year of their 
creation although  the low energy solutions have names that end with  40. 
Clearly that database has grown rapidly in the last two decades.
While the {\em pp} data now extends to 3 GeV, the {\em np} data are
limited to 1.3 GeV.  Surprisingly, the solutions from
SM97 to WI00 remain very closely the same and are very stable
with regard to new data. We have  used the solutions SP00, FA00 and 
WI00 in our calculations and found results that differ but marginally. 
Thus hereafter in the main  we refer solely to the results of calculations
 based upon SP00. 
The findings are equally valid for other more recently dated solutions.
In our practical applications however when  new potentials are 
sought their
generation is based upon the most current solution \cite{SAID}.

The VPI/GWU solutions \cite{Arn82} are parameterizations of 
the elastic channel NN S-matrix. They consider 
\begin{equation}
S_1=(1+iK_4)(1-iK_4)^{-1}
\end{equation}
which  inverts to give
\begin{equation}
K_4=i(1-S_1)(1+S_1)^{-1}=Re\, K_4+iIm\, K_4.
\end{equation}
The real part of this K-matrix is related to a unitary S-matrix ($S_6$)  
and therewith phase shifts $\delta^\pm$ and $\epsilon$ are defined by
\begin{equation}\label{eqn_II.1}
S_6={(1+i\,Re\, K_4)\over (1-i\, Re\, K_4)}=
\left\{ \begin{array}{cc} 
\cos 2\varepsilon \exp {2i\delta^-} & 
i \sin 2\varepsilon \exp {i(\delta^-+\delta^+)}\\
i \sin 2\varepsilon \exp {i(\delta^-+\delta^+)}&  
\cos 2\varepsilon \exp {2i\delta^+}
\end{array}\right\}.
\end{equation}
The absorption parameters $\rho^\pm$ and $\mu$ 
 relate to the imaginary part of that K-matrix by
\begin{equation}\label{eqn_II.2}
Im\, K_4=\left\{
\begin{array}{cc}
\tan^2 \rho^- & 
\tan\rho^- \tan\rho^+\cos \mu\\
\tan\rho^- \tan\rho^+\cos \mu &
\tan^2 \rho^+
\end{array}
\right\}.
\end{equation}
These relations simplify to $K=\tan\delta+i\tan^2\rho$ for 
uncoupled channels.

In our study, real NN potentials derived from fixed angular 
momentum inverse
scattering theory have been used. They have been generated from 
inversion algorithms predicated upon the Gel'fand--Levitan--Marchenko
integral equations which  physically link to the radial 
Schr\"odinger equation
of a fixed angular momentum,
\begin{equation}  \label{eqn_II.inv1}
\left[ -{d^2 \over dr^2} + {\ell ( \ell+1) \over r^2 } + 
{2 \mu \over \hbar^2} V_\ell (r)  
\right] \psi_\ell (r,k) = k^2 \psi_\ell (r,k),
\end{equation}
where $V_\ell (r)$ is a local and energy independent operator in
coordinate space. Substituting
\begin{equation}  \label{eqn_II.inv2}
q(r)={\ell (\ell +1)\over r^2} + {2\mu\over\hbar^2} V_\ell
(r),\quad\mbox{and}\qquad \lambda=k^2,
\end{equation}
identifies Eq.\,(\ref{eqn_II.inv1}) as a  Sturm--Liouville equation
\begin{equation}  \label{eqn_II.inv3}
\left[ - {d^2 \over dx^2} + q(x) \right] y(x) = \lambda y(x).
\end{equation}

There are two equivalent inversion algorithms for the Sturm--Liouville
equation, which one identifies as the Marchenko and the 
Gel'fand--Levitan inversion. 
Both yield principally the same solution and  numerically 
they are  complementary. 
The salient features are outlined for the case of uncoupled channels.  
For coupled channels the inversion equations are 
matrix equations with input and translation kernels 
correspondingly generalized.

In the Marchenko inversion the experimental information enters via the
$S$-matrix, $S_{\ell}(k)=\exp(2i\delta_{\ell}(k))$, with which  
an input kernel is defined 
in the form of a Fourier-Hankel transform
\begin{equation} \label{eqn_II.inv4}
F_\ell (r,t) = -\frac{1}{2\pi} \int_{-\infty}^{+\infty} h^+_\ell(rk)
\left[ S_\ell(k)-1 \right] h^+_\ell(tk) dk,
\end{equation} 
where $h^+_\ell(x)$ are  Riccati-Hankel functions. 
This input kernel when used in the Marchenko equation,  
\begin{equation} \label{eqn_II.inv5}
A_\ell (r,t)+F_\ell (r,t)+\int_{r}^{\infty}A_\ell(r,s)F_\ell(s,t)ds = 0,
\end{equation}
specifies the translation kernel
$A_{\ell}(r,t)$. The potential of Eq.\,(\ref{eqn_II.inv1})
is a boundary condition for that translational
kernel,
\begin{equation} \label{eqn_II.inv6}
V_{\ell}(r)=-2\frac{d}{dr}A_{\ell}(r,r).
\end{equation}

The Gel'fand--Levitan inversion requires not 
the $S$-matrix but rather the Jost-function as spectral input. 
The latter is related to the $S$-matrix by
\begin{equation} \label{eqn_II.inv7}
S_{\ell}(k)=\frac{F_{\ell}(-k)}{F_{\ell}(k)}.
\end{equation}
The Gel'fand--Levitan input kernel then is defined as the  
Fourier-Bessel transform 
\begin{equation} \label{eqn_II.inv8}
G_\ell (r,t) = \frac{2}{\pi} \int_{0}^{\infty} j_\ell (rk) \left[
\frac{1}{|F_\ell (k)|^2}- 1  \right] j_\ell (tk) dk,
\end{equation}
where $j_{\ell}(x)$ are Riccati-Bessel functions. 
The Gel'fand--Levitan integral equation
\begin{equation} \label{eqn_II.inv9}
K_\ell (r,t)+G_\ell (r,t)+\int_{0}^{r}K_\ell (r,s)G_\ell (s,t)ds = 0,
\end{equation}
also defines a translational  kernel with  boundary condition 
\begin{equation} \label{eqn_II.inv10}
V_{\ell}(r)=2\frac{d}{dr}K_{\ell}(r,r).
\end{equation}
The boundary conditions Eq.\,(\ref{eqn_II.inv6}) and (\ref{eqn_II.inv10})
yield identical potentials. 

\begin{figure}[tbp]
\centering
\epsfig{file=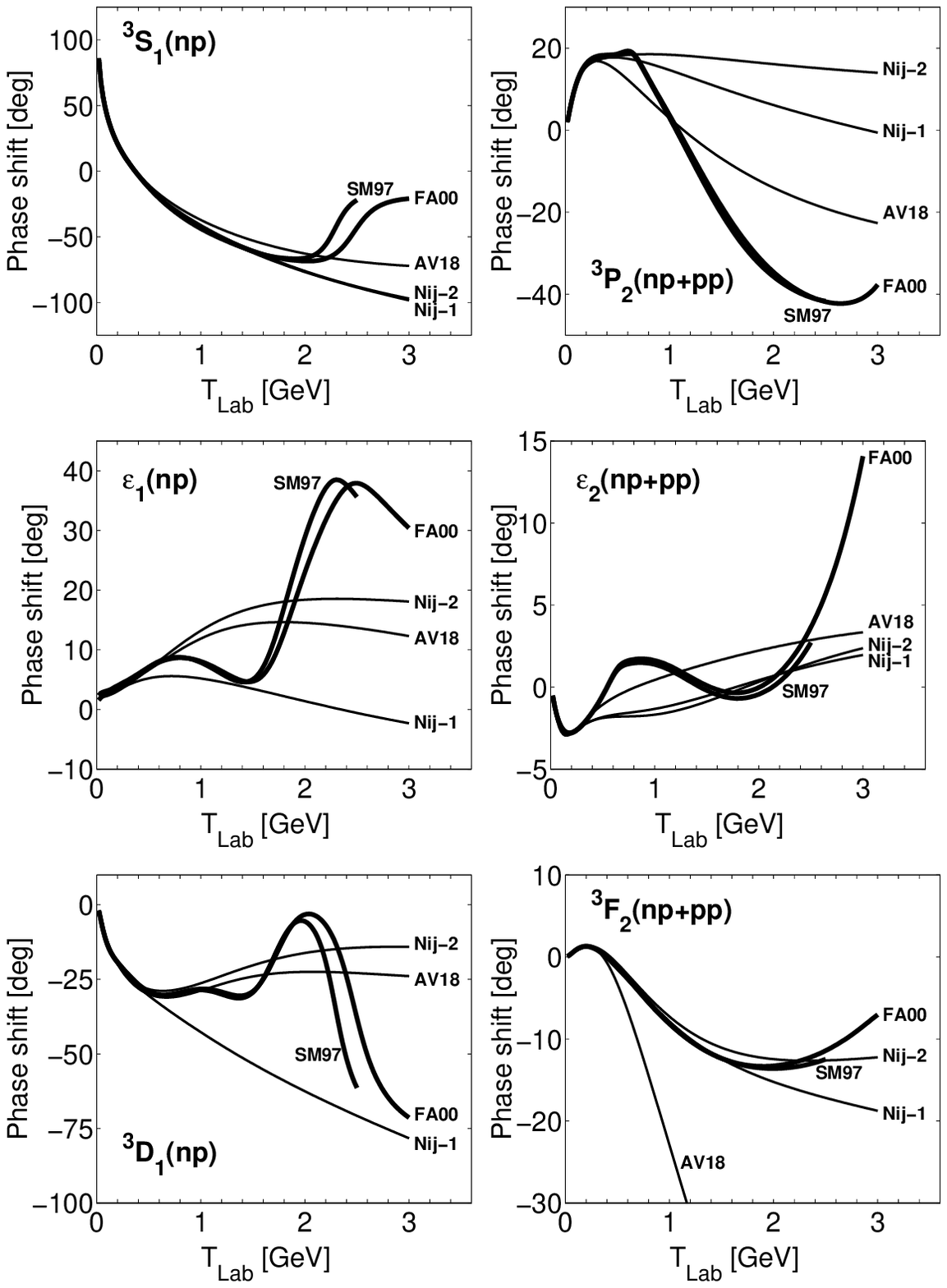,scale=.85}
\caption{Coupled channel phase shifts  for SM97 $T_{Lab}<2.5$ GeV, 
FA00 $T_{Lab}<3$ GeV,
and reference phase shifts using 
Nijmegen(Nij-1, Nij-2) and Argonne (AV18) potentials.}
\label{figure_II.2}
\end{figure}

Determination of the input kernels from data,  phase shift 
functions $\delta(T_{Lab}(k))$, 
or K-matrices $K(T_{Lab}(k))$, requires  an accurate  interpolation and
extrapolation of that data. In all practical applications 
rational functions are very appropriate. In this work we 
made a representation of
data  for $T_{Lab}(k)\leq 3$ GeV where the order $N$ of 
the rational functions
$P^{[2N-1,2N]}(k)$ was chosen to be as small as possible. 
Typically $2<N<6$. An
implication is that extrapolations of $\delta(k)$ from 
the highest energy (last)
data point $k_{max}$ to infinity  do not change sign  and 
$\lim_{k\to \infty}\delta(k)\sim 1/k$. We control the 
rational function fit with 
weight functions which guarantee that those fits will be particularly 
accurate for some desired interval and less stringent elsewhere. 
For example, the channels $^1S_0,\ ^1P_1,\ ^3P_{0,1},\ ^3D_2$ and $^1F_3$ 
were weighted with $w_{Low}=1$ for 
$T_{Lab}< 1.2$ GeV and for larger energies,  $w_{High}= 0.05$. 
For the $^1D_2$ and $^3F_3$ channels, the
cut between $w_{Low}$ and $w_{High}$ was 300 MeV. Consequently 
the rational
functions used in the inversion algorithm ensure that the resulting
potentials will give the desired values of phase shifts from 
solutions of the 
\begin{figure}[tbp]
\centering
\epsfig{file=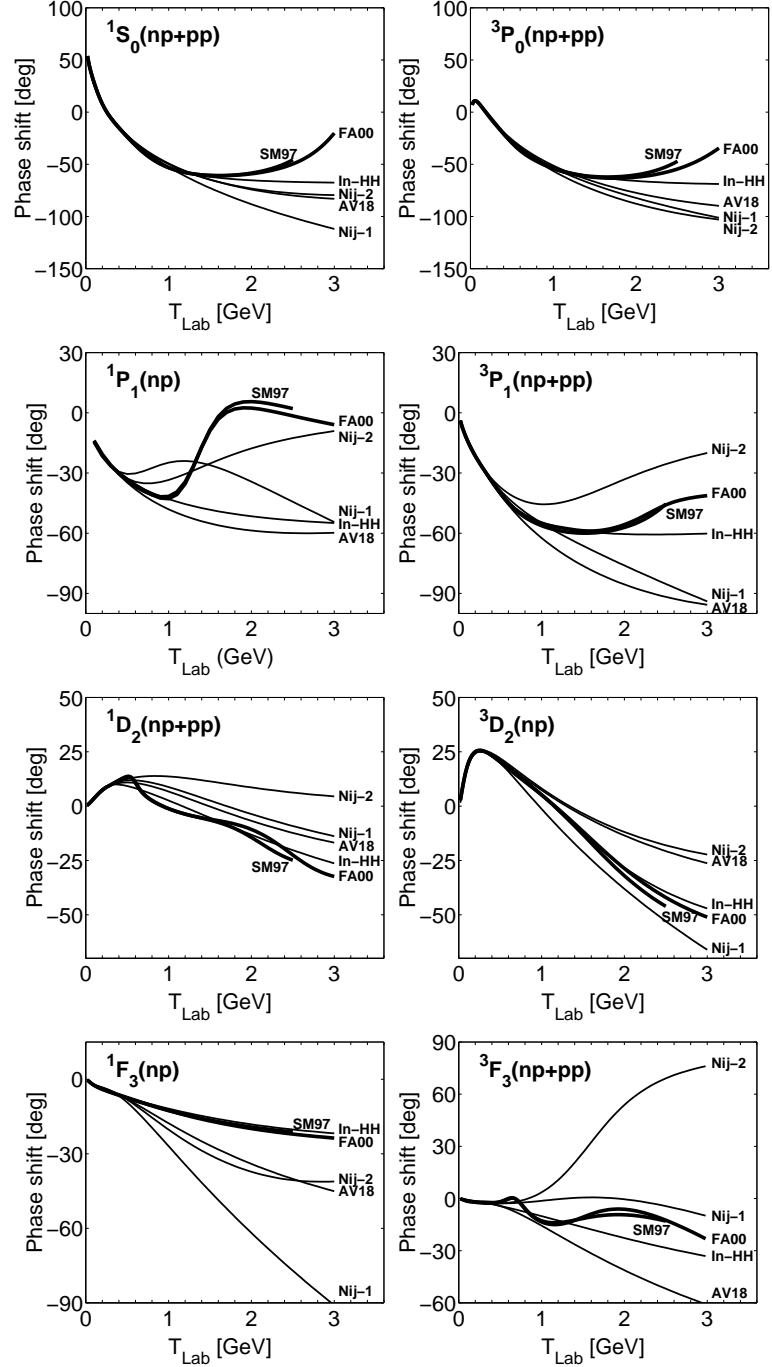,scale=0.85}
\caption{Single channel phase shifts for SM97 $T_{Lab}<2.5$ GeV), 
FA00 $T_{Lab}<3$ GeV, and reference phase shifts using
inversion (In-HH), Nijmegen (Nij-1, Nij-2) and Argonne (AV18) potentials.}
\label{figure_II.1}
\end{figure}
Schr\"odinger equation. Such is evident from the comparisons given in 
Figs.\,\ref{figure_II.1} and \ref{figure_II.2}. 
Therein 
the fits to the phase shifts to 300 MeV resulting from all three models 
are considered as high quality. 
Single and coupled channel phase shifts from 
SM97 and FA00 solutions for $J\leq 3$  are shown together with 
values  found from calculations made using three  potential models. 
These model
phase shifts were generated with Nijmegen-I and Nijmegen-II 
\cite{Sto93},
and Argonne AV18 \cite{Wir95} interactions,  and with potentials 
determined using 
Gel'fand--Levitan--Marchenko inversion\cite{Ger00,Kir89,San97}. 

\begin{figure}[tbp]
\centering
\epsfig{file=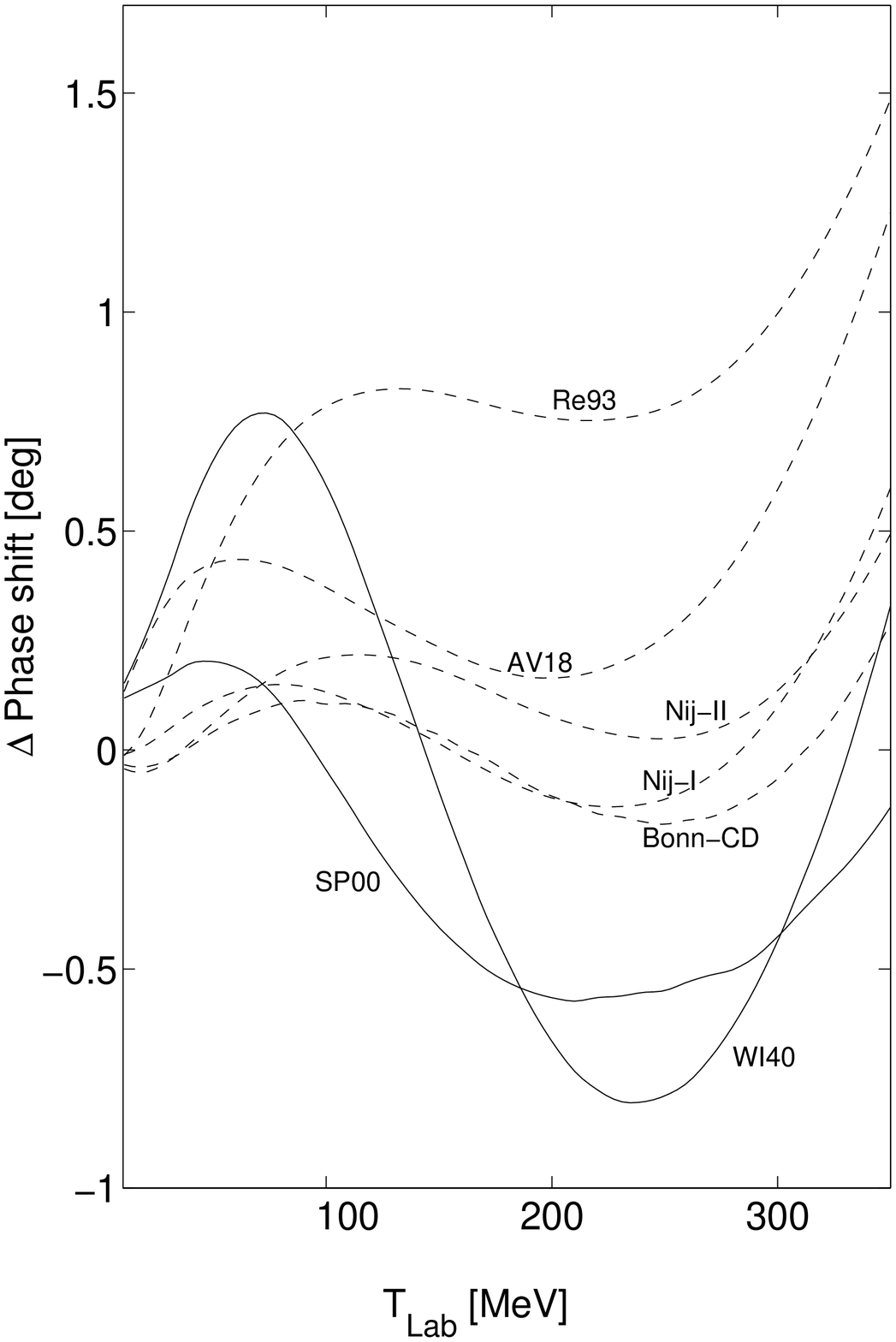,scale=0.6}
\caption{{\em np} $^3P_0$ phase shift differences with respect to
  Nijmegen PWA93.}
\label{figure_II.4}
\end{figure}

On the scale to 3 GeV the  OBE model results clearly diverge 
from data.  
As with the phase shift analyses, OBE 
potentials have received several
critical reviews\cite{Mac01a,Mac01}, including observations 
that there are small variations between phase shift analyses 
and potential model results  in the subthreshold
domain $T_{Lab} <300$ MeV \cite{Had01}. A theoretically 
stable result
would require many quantities, that need be specified {\em a priori}, 
to be  determined from other sources. At present that does 
not seem feasible and all current potentials rely upon 
fits of many of their 
parameters to the same data. All such fits, however, have been 
made independently of each other and are
based upon differing theoretical specifications of 
the boson exchange model dynamics.   
\begin{figure}[tbp]
\centering
\epsfig{file=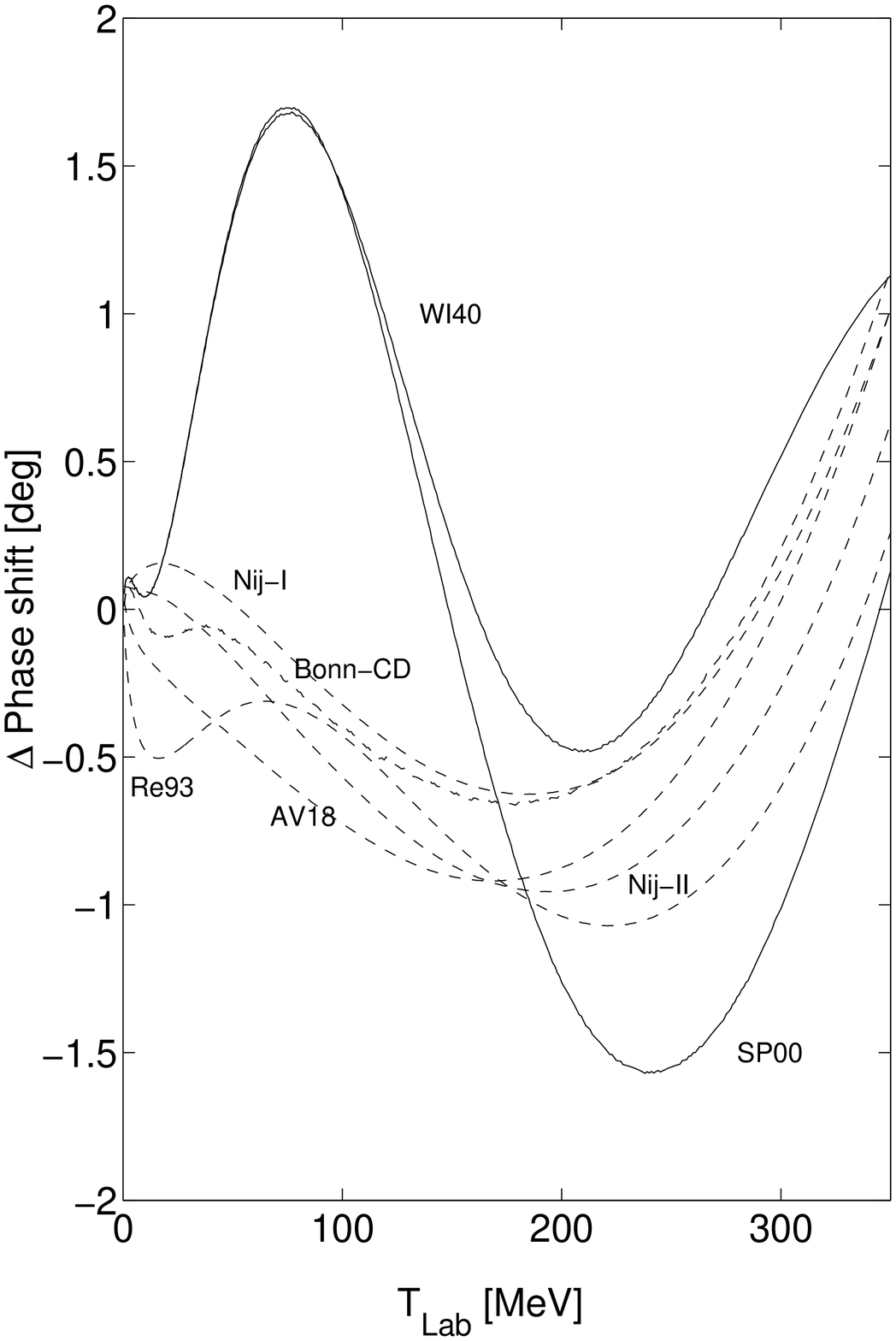,scale=0.6}
\caption{{\em np} $^1S_0$ phase shift differences with respect to
  Nijmegen PWA93.}
\label{figure_II.3}
\end{figure}   
In Figs.\,\ref{figure_II.3} and \ref{figure_II.4} 
we give a quantitative demonstration of the ensuing differences. 
Therein the Nijmegen phase shift analysis PWA93 \cite{PWA93}
has been  used as reference values for various other phase shift 
solutions and potential
predictions for the {\em np} $^1S_0$, and $^3P_0$ channels.
Such differences are characteristic of variations between finite
power series expansions of data in a finite interval. A mathematical
property of such finite power series expansions within an 
interval is that, while the data in the interval will be well
reproduced, continuations beyond that interval can radically
diverge. Such a property is in evidence in Figs.\,\ref{figure_II.1} 
and \ref{figure_II.2}, and  
that variance is the reason for the caveat often espoused  that use of
OBEP beyond the fitted  energy range should be prohibited. 
Be that as it may, one could expect from a consistent theory 
that such extrapolations, albeit in error, would be the same. 
Clearly they are not. However of one thing we can be sure, the lack of 
physics  with these models lies within the interaction 
distance $<1$ fm. The optical model approach we present 
is an attempt within the frame of potential scattering theory 
to account for and identify such short range properties.  

It is apropos to make a brief remark on the long range character of the
NN potential which theoretically is identified with OPEP.
In the phase shift analysis PWA93 by the Nijmegen group and 
in that of Bonn-CD-2000 of Machleidt, such  character is enforced.
Indeed that precise character re-emerges when either of those phase
shift functions are used as input to a Gel'fand--Levitan--Marchenko
inversion. On the other hand, the VPI/GWU group makes no use
of OPEP in any of their solutions. Exactly the same quantum
inversion of the SM94 solution does not give OPEP except on average
which might be interpreted as signaling the importance of
nonlocality\cite{San97}.   

\begin{figure}[htb]
\centering
\epsfig{file=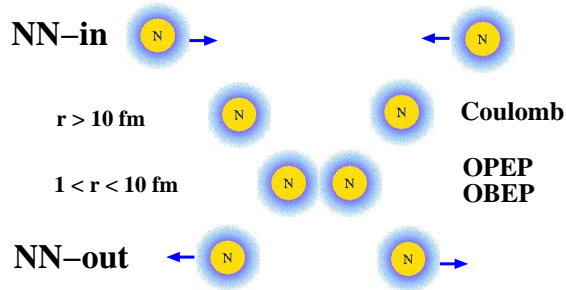,scale=.5}
\caption{Interaction scheme for low energy scattering, 
$0< T_{Lab} <300$ MeV}
\label{figure_II.5}
\end{figure}

\begin{figure}[tbh]
\centering
\epsfig{file=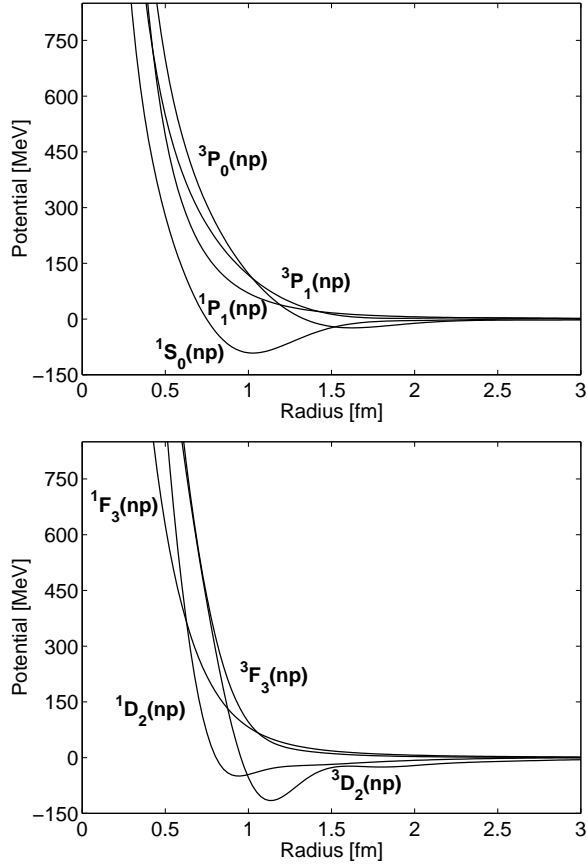,scale=0.8}
\caption{Nucleon-nucleon inversion potentials using SP00 phases}
\label{figure_II.6}
\end{figure}

\begin{figure}[htb]
\centering
\epsfig{file=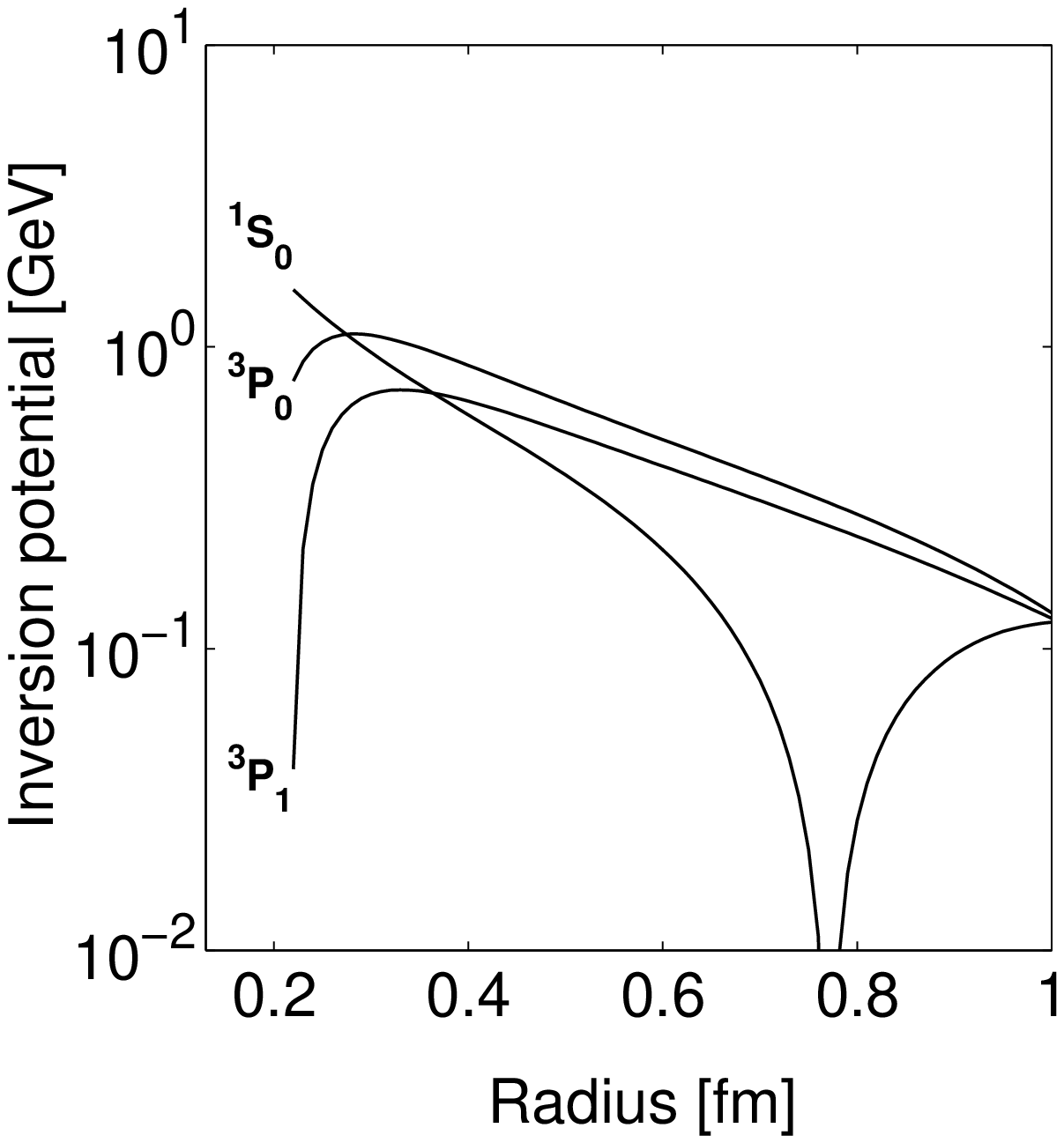,scale=0.8}
\caption{{\em np} $^1S_0$ and $^3P_{0,1}$ inversion potentials 
using SP00  real phase shift solution to 3 GeV.}
\label{figure_II.7}
\end{figure}

Despite limitations as discussed above, the OBEP remain the best 
motivated potential models for low energy scattering. They do yield
high quality fits to the phase shifts in that domain. Such is useful
for us in our quest to interpret data with increasing energy.
In Fig.\,\ref{figure_II.5}
 we show an interaction scheme 
in terms of radial separation that is suitable for low energy scattering.  
This scheme is supported by potentials determined by inversion 
which reproduce the low energy  
phase shifts used as input  to an accuracy 
$|\delta(exp.)-\delta(rat.)|<0.25$ degrees. 
Such inversion potentials have been made also to 
follow closely the SP00 real
phase shifts to 3 GeV and these are shown in 
Fig.\,\ref{figure_II.6}.
They possess 
a long range Yukawa tail, a medium range attraction $\sim$1-2 fm and 
a strong short range repulsion with an onset at 1 fm.
These potentials are energy independent so that the long and
medium range potential properties diminish in 
importance for kinetic  energies
above 500 MeV. For projectiles with $T_{Lab}>1.5$ GeV essentially
only the repulsive core of these potentials remains of 
significance for scattering. 
Thus inversion potentials have also been obtained with the 
SP00 real phase shifts
to 3 GeV  using $w_{Low}=0.1$ for $T_{Lab}<1.2$ GeV  
and $w_{High}=1$ for higher energies, to  emphasize the high energy
data and fix more stringently the short range ($<1$ fm) 
character of the deduced
interaction. The short range properties of inversion potentials 
so found  are 
displayed  in Fig.\,\ref{figure_II.7}.
Clearly, the $^1S_0$ and $^3P_{0,1}$ 
inversion potentials based upon SP00 
real phase shifts which  extend to 3 GeV are soft core interactions.  
We neglected in this analyses the $np$ $^1P_1$ channel due to 
the limited data set for  $T_{Lab}<1.2$ GeV.  The higher 
partial waves are strongly 
screened by the centripetal barrier and so also are not considered here.
The core strengths of these $^1S_0$ and $^3P_{0,1}$
potentials reach a shoulder and maximum with a 
typical value $\sim1$ GeV at a radius of 0.3 to 0.4 fm. 
It is worth noting that the shoulder/maximum aspect of the core is a
result of flat minima  between 1.5 and 2 GeV in the $^1S_0$ and $^3P_{0,1}$  
SP00 phase shift functions. For higher partial waves,   phase shift minima 
lie beyond 3 GeV.  As the experimental
phase shifts are limited to 3 GeV we have confidence in the 
specified inversion potentials
only to about 0.25 fm. The shorter distance values reflect only our
extrapolation of these phase shifts being 
$\lim_{k\to \infty}\delta(k)\sim 1/k$.

\begin{figure}[htb]
\centering
\epsfig{file=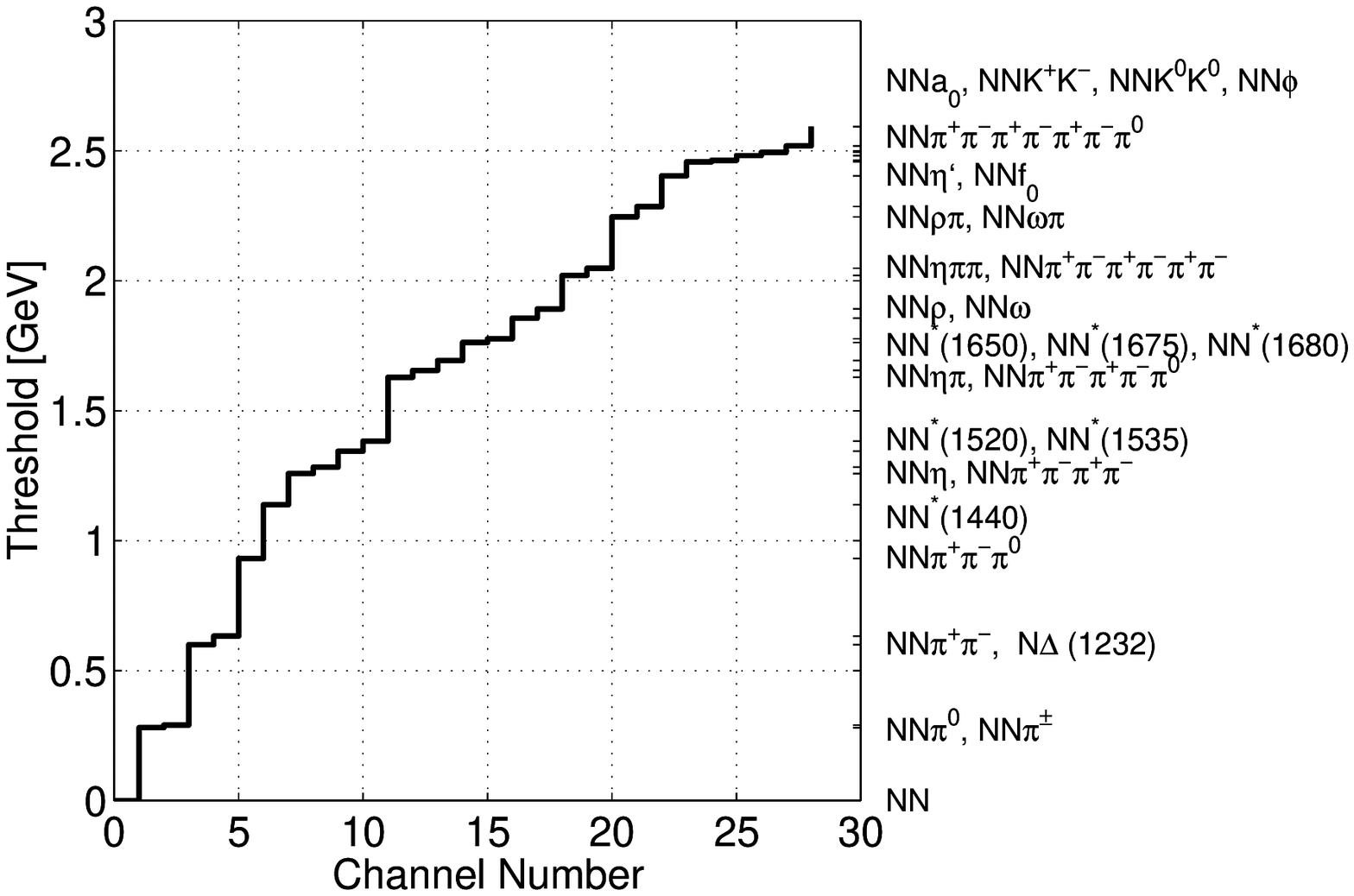,scale=.8}
\caption{Thresholds for production processes in NN scattering.}
\label{figure_II.8}
\end{figure}

Above 300 MeV reaction channels open and 
the elastic channel S-matrix no longer is unitary. 
In Fig.\,\ref{figure_II.8}
we show the
gradual increase of the open  channels in NN scattering
which includes resonances as well as single and multiple
production thresholds. Only the $\Delta$(1232) resonance has a
low energy threshold and a relative small width of 120 MeV. 
Therefore it is the
only resonance we expect to be obviously visible in the energy variation of
the elastic scattering phase shifts. In particular one 
notices typical variations
in the $^1D_2$, $^3F_3$, and $^3PF_2$ channels. Otherwise the phase shifts 
to 3 GeV are very smooth slowly changing functions of energy
in all channels. Such is a condition for the suitability of a 
potential model of scattering governed by {\em quasi} macroscopic
geometric entities. In nucleon-nucleus (NA) scattering,  
entities of that ilk 
are epitomized by the parameters of Woods--Saxon potentials.
For the NN case, we have used previously \cite{Ger98} a 
local Gaussian in this same manner,
noticing therefrom spin-isospin coupling effects more
substantial than found with NA scattering. It is also worth noting that the
absorption in those NN optical potentials for this energy range were
not at the  geometric limit of a fully absorptive disc.
Together with the strong spin-isospin coupling, this property infers
optical potentials that are strongly channel dependent in
contrast to the NA case for which assumed central and spin-orbit
potentials are partial wave independent.

The plethora of reaction channels that open  to 3 GeV, and the requirement
of an NN optical potential prescription discussed above, mean that it is 
an interesting but very challenging task for any microscopic model
to uniquely link QCD substructures to NN scattering 
phase shift functions in analogy to that successful prescription by 
which NA optical potentials have
been determined by folding effective NN interactions with nuclear density
matrix elements\cite{Amo00}. Given the success of optical models
for NA scattering and as the  distorted 
wave Born approximation (DWBA) is  well known and 
successful in nuclear, atomic
and molecular physics to study inelastic reactions, 
it is intriguing to conjecture
use of the DWBA to analyze inelastic reaction channels of 
NN induced reactions.

\begin{figure}[htb]
\centering
\epsfig{file=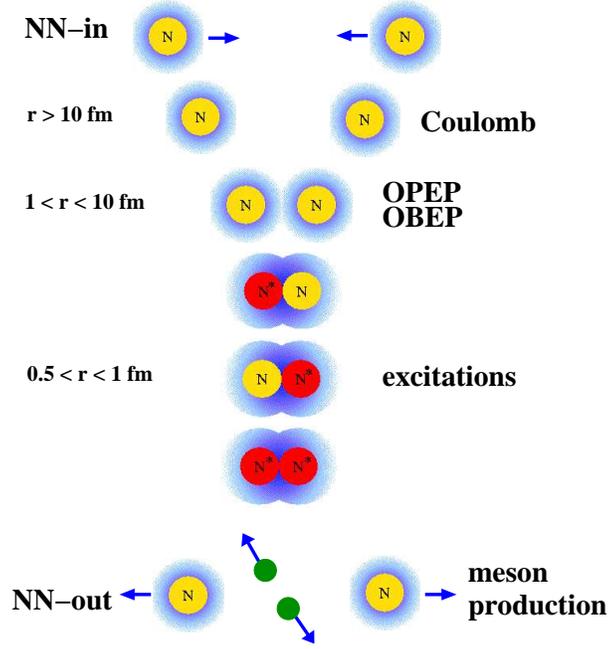,scale=.5}
\caption{Interaction scheme for medium energy scattering, 
$0.3 <  T_{Lab} < 2$ GeV.}
\label{figure_II.9}
\end{figure}

In the spirit of visualization of NN scattering 
shown in Fig.\,\ref{figure_II.5}, we
now include the importance of the reactive and resonant 
content pictorially
in Fig.\,\ref{figure_II.9}.
This we consider relevant 
for $0.3<T_{Lab}<2$ GeV. The upper 
limit is significant here as we discuss later, but for now it suffices
that the potential shoulder and maximum seen in Fig.\,\ref{figure_II.7} 
are $\sim$1 GeV. Now we identify
some specifics in the $0.5 < r <1$ fm range. We conjecture that the 
two colliding hadrons are retained in hadronic states 
throughout the process. 
We allow one of the two nucleons to be excited, say into
a $\Delta$(1232), while the other remains in the ground state. The 
excitation may be exchanged between the two hadrons as well, and both
nucleons may be excited to an intermediate resonant state. The
production of mesons then can only occur from one or both of the
two separate QCD entities. The essential feature is that 
in the energy range,
the predominant scattering processes are those retaining identifiable 
hadronic entities.
Within an optical potential representation,
attendant flux loss equates to a diffuse absorption extending
radially to 3 fm and possibly more. The bulk of such absorption
however lies significantly within  1 fm.

\begin{figure}[htb]
\centering
\epsfig{file=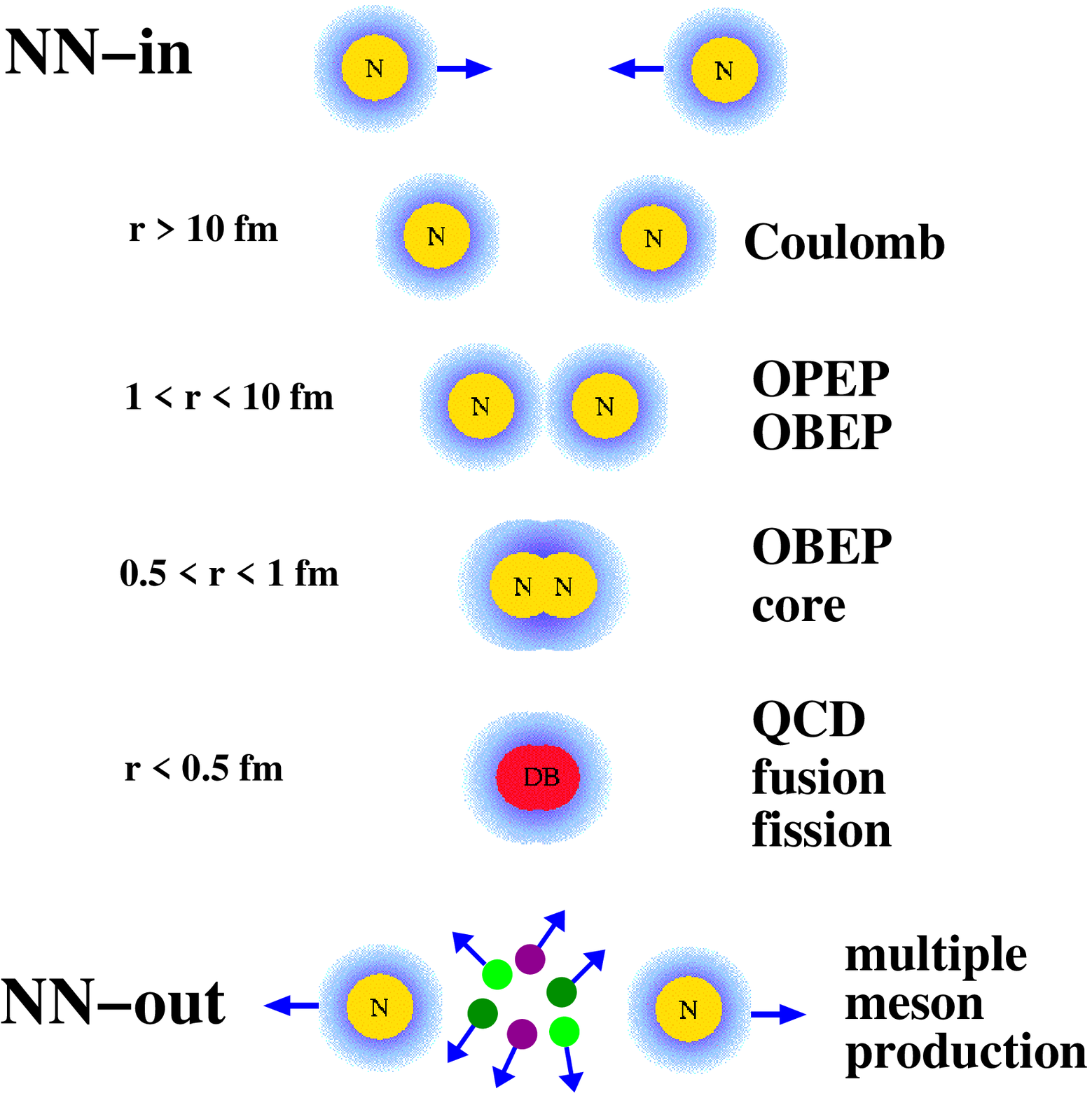,scale=.5}
\caption{Interaction scheme for high energy scattering, $T_{Lab}>2$ GeV}
\label{figure_II.10}
\end{figure}

It requires 2 GeV and more for the two nucleons to surpass 
the core potential
barrier and to fuse into a compound system. 
This is visualized with the scattering 
sequences shown in Fig.\,\ref{figure_II.10}.
An objective of our optical model studies is to
substantiate this conjecture of fusion and fission of resultant 
compound dibaryonic systems dominating the scattering for 
this energy regime.

To describe this developing system for $300$ MeV $<T_{Lab}<3$ GeV
we will use Feshbach theory to specify the
optical potential\cite{Fes62}. An important feature of 
that theory is the projection
operator formalism with $P$  and  $Q$  subspaces, which 
divide the complete
Hilbert space $(P+Q)=1$, into the elastic scattering channel, 
the $P$ space, and
all inelastic and reaction channels which are  contained in $Q$ space. 
This theory then assumes a hierarchy of complication 
in $Q$ space of which 
doorway states are the simplest. Doorway states are 
characterized to be the only
means to leave and to return to the elastic channel. 
Each doorway state in
this approach infers a complex and separable component in the optical
potential with an energy dependent strength. If a very large number of
doorway states contribute, the effect equates to 
a local potential operator.
This was the basis of our previous study \cite{Ger98}.

\subsection{Formal Potential Model}

It is generally accepted that a valid covariant description
of NN scattering formally is given
by the Bethe--Salpeter equation
\begin{equation} \label{eqn_III.3}
{\cal M} = {\cal V} + {\cal V}{\cal G}{\cal M}\ ,
\end{equation}
where  $\cal M$ are invariant amplitudes
that are based upon all connected two particle irreducible diagrams.
This equation serves generally as an ansatz for
approximations. Of those, the three dimensional reductions
are of great use which allow the definition of a potential
\cite{Bla66,Gro82}. 
In particular,
the Blankenbecler--Sugar reduction \cite{Bla66} 
gives an equation very often used  for 
applications with NN scattering \cite{Mac01,Par70}.
This reduction is obtained from Eq.\,(\ref{eqn_III.3}), which 
in terms of four-momenta is
\begin{equation}\label{eqn_III.4}
{\cal M} ( q^\prime,q;P ) = {\cal V} ( q^\prime,q;P ) + 
\int d^4k\ {\cal V} ( q^\prime,k;P )
\  {\cal G} (k;P)\ {\cal M} ( k,q;P),
\end{equation}
where  the  propagator
\begin{equation}\label{eqn_III.5}
{\cal G} (k;P) = { i \over (2 \pi)^4} \left[ { \frac12\fdag P
+ \fdag k + M \over (\frac12 P + k)^2 - M^2 + i \varepsilon}\right]_{(1)}
\left[ { \frac12\fdag P
- \fdag k + M \over (\frac12 P - k)^2 - M^2 + i \varepsilon}\right]_{(2)}.
\end{equation}
The subscripts
refer to  nucleon (1) and (2) respectively. 
In the CM system $P = (\sqrt{s},0)$,
which is just the  total energy $E=\sqrt{s}$. 
In particular, the Blankenbecler--Sugar reduction 
of the propagator $\cal G$
uses the covariant form
\begin{equation}\label{eqn_III.6}
{\cal G}_{\rm BS} (k,s) = - {\delta (k_0) 
\over (2 \pi)^3 } { M^2 \over E_k}
{\Lambda^+_{(1)} ({\bf k}) \Lambda^+_{(2)} (-{\bf k}) \over
\frac14s - E_k^2 + i \varepsilon}\ ,
\end{equation}
with  positive energy projectors
\begin{equation}\label{eqn_III.7}
\Lambda^+_{(i)} ({\bf k}) = \left( { \gamma^0 E_k - \vec{\gamma} \cdot
{\bf k} + M \over 2M} \right)_{(i)}.
\end{equation}
The amplitudes are now expressed with the reduced terms
and they  satisfy a three-dimensional equation
\begin{equation}\label{eqn_III.8}
{\cal M} ({\bf q}^\prime,{\bf q}) = {\cal V} ({\bf q}^\prime,{\bf q}) +
\int { d^3k \over (2 \pi)^3 }{\cal V} ({\bf q}^\prime,{\bf k})
 { M^2 \over E_k} {\Lambda^+_{(1)} ({\bf k}) 
\Lambda^+_{(2)} (-{\bf k}) \over
{\bf q}^2 - {\bf k}^2 + i \varepsilon}
{\cal M} ({\bf k},{\bf q}).
\end{equation}
Taking matrix elements with only positive energy spinors, 
an equation with  minimum relativity results
for the  NN $t$-matrix, namely
\begin{equation}\label{eqn_III.9}
{\cal T} ({\bf q}^\prime,{\bf q}) =  {\cal V} ({\bf q}^\prime,{\bf q}) +
\int { d^3k \over (2 \pi)^3}  {\cal V} ({\bf q}^\prime,{\bf k})
 { M^2 \over E_k} {1 \over {\bf q}^2 - {\bf k}^2 + i \varepsilon}
{\cal T} ({\bf k},{\bf q}).
\end{equation}
Using  the substitutions
\begin{equation}\label{eqn_III.10}
T ({\bf q}^\prime,{\bf q}) = \left( M \over E_{q^\prime} 
\right)^{\frac12} 
{\cal T} ({\bf q}^\prime,{\bf q})
\left( M \over E_{q} \right)^{\frac12}
\end{equation}
and
\begin{equation}\label{eqn_III.11}
V ({\bf q}^\prime,{\bf q})
= \left( M \over E_{q^\prime} \right)^{\frac12}
{\cal V} ({\bf q}^\prime,{\bf q})
\left( M \over E_{q} \right)^{\frac12},
\end{equation}
a simplified form of the t-matrix is obtained. It is the familiar 
Lippmann--Schwinger equation  
\begin{equation}\label{eqn_III.12}
 T ({\bf q}^\prime,{\bf q}) =  V ({\bf q}^\prime,{\bf q}) +
\int { d^3k \over (2 \pi)^3}  V ({\bf q}^\prime,{\bf k})
 {M \over {\bf q}^2 - {\bf k}^2 + i \varepsilon}
 T ({\bf k},{\bf q})\ .
\end{equation}

Of use is an  equivalent Lippmann--Schwinger  equation 
for the  wave function. Formally, this equivalence 
is proven with  the M\o ller distortion operator which relates
the  free wave function with the scattered wave
and uses   the relation between scattering
amplitude and potential, $T^{(\pm)}\Phi=V\Omega^{(\pm)}\Phi$. 
Finally, we use the equivalence between the Lippmann--Schwinger
integral equation  and the Schr\"odinger equation so that 
\begin{equation} \label{eqn_III.13}
\left(-\Delta +{M\over \hbar^2} V({\bf r})-k^2\right) \psi(\bf{r,k}) = 0.
\end{equation}
When we identify the potential scale $M$  with the 
two particle reduced mass
\begin{equation}\label{eqn_III.14}
M=2\mu =2{m_1m_2\over m_1+m_2}
\end{equation}
we guarantee consistency with the low energy limit of the 
Schr\"odinger equation and use, therein, of 
NN OBE  reference potentials. However, 
a careful and consistent treatment of the  
$M/E$ factors in Eqs.\,(\ref{eqn_III.10}) and
(\ref{eqn_III.11}) is necessary whenever it is 
important to take relativity into account.
Minimal relativity enters  in the calculation of $k^2$ by 
\begin{equation}\label{eqn_III.15}
s=(m_1+m_2)^2+2m_2T_{Lab}=
\left(\sqrt{k^2+m_1^2}+ \sqrt{k^2+m_2^2}\right)^2\ ,
\end{equation}
where
\begin{equation}\label{eqn_III.16}
k^2 = { \displaystyle m_2^2 ( T_{Lab}^2 + 2 m_1 T_{Lab}) \over
\displaystyle ( m_1 + m_2 )^2 + 2 m_2 T_{Lab}}\ .
\end{equation} 
For equal masses this reduces to $ k^2 =  s/4- m^2$.

\section{An algorithm for the optical and boundary condition models}

We distinguish between  three Hamiltonians. They are
the  {\em reference}  Hamiltonian  $H_0$, a projected Hamiltonian $H_{PP}$,
and a full optical model Hamiltonian $\cal H$. The first of these, the 
{\em reference} Hamiltonian $H_0:=T+V_0$, invokes a given potential $V_0$ 
for which one can find Schr\"odinger equation reference  
solutions. The physical outgoing solutions
$\psi_0:=\psi_0^+({\bf r,k},E)$ of $H_0$ we suppose gives a 
unitary  S-matrix. 
We assume further that this Hamiltonian is completely 
specified such that
evaluation of any quantity, wave function, S-matrix, 
K-matrix {\em etc.} 
is facilitated. The Feshbach projection 
operator formalism\cite{Fes62} is used to give 
the {\em projected}  Hamiltonian,
$P H_0P=H_{PP}$, derived from $H_0$.  We presuppose 
completeness, $P+Q=1$, 
and a finite rank representation of the Q space
\begin {equation} \label{eqn_III.17}
Q:=\sum_{i=1}^N |\Phi_i><\Phi_i|=\sum_{i=1}^N |i><i|,
\end{equation}
with the Q space basis functions $|\Phi_i>$ interpreted 
as doorway states. With these doorway states we make the link 
between  the  QCD and  the hadronic sectors; the latter  
encompassing
nucleons, mesons and other free particles.
Thus we will  assume that meson creation/annihilation 
occurs only in the highly nonlinear 
QCD sector so that   Q space wave functions are  projections
of such  processes onto hadronic particle coordinates. 
The third of our
Hamiltonians, the {\em full optical model} Hamiltonian,
comprises the reference Hamiltonian
$H_0$ and the proper optical model potential $\cal V$. 
That potential  
is  complex and nonlocal, {\em viz.} separable of finite rank,  
${\cal H}:=T+V_0+{\cal V}(r,r';lsj,E)$. 
Separable potentials are popular  representations of 
NN potentials 
which are designed to serve quite a wide range of 
purposes \cite{Kwo97}.

The Schr\"odinger equation specified with $\cal H$ 
has regular physical solutions  
$\Psi^+:=\Psi^+({\bf  r, k},E)$ whose  asymptotic 
boundary conditions 
we deem to match with the {\em experimental} 
elastic channel S-matrix. 
Specifically, for these experimental S-matrices 
we have used  the continuous  
solutions SP00 from  VPI/GWU \cite{SAID}.
The reference potential $V_0$ and separable potential
form factors are to be specified in detail 
with any application.

\subsection{Towards a full optical potential model}

To obtain the optical potential on the basis 
of a given reference potential,
we express first the solutions of the projected 
Hamiltonian
in terms of the reference Hamiltonian and 
the {\em a priori} defined 
Q space projector. 
The Schr\"odinger equation $(E-H_0)|\psi_0>=0$ and 
its solutions are used to
express the solutions of
$(E-H_{PP})|\psi_P>=0$. The latter is equivalent to 
the Schr\"odinger equation
\begin{equation}\label{eqn_III.18}
(E-H_{PP}-H_{QP}-H_{PQ}-H_{QQ})
|\psi_P>=-H_{QP}|\psi_P>
\end{equation}
and the Lippmann--Schwinger equation 
\begin{eqnarray} \label{eqn_III.19}
|\psi_P>&=&|\psi_0>-{1\over(E^+-H_0)}H_{QP}|\psi_P>
\nonumber \\ &=&|\psi_0>-\sum_jG^+|j><j|H_{QP}|\psi_P>.
\end{eqnarray}

These equations are still very general and do not 
depend upon a specific
representation. However, in the following we assume 
a partial wave expansion
in terms of spherical harmonics, spin and isospin 
state vectors and radial functions.  
The following equations are identified as radial equations
with the set of quantum numbers suppressed.    

Projector orthogonality $PQ=QP=0$ implies that 
\begin{equation} \label{eqn_III.20}
0=<i|\psi_P>=<i|\psi_0>-<i|G^+ H_{QP}|\psi_P>,
\end{equation}
and thus
\begin{equation}\label{eqn_III.21}
<j|H_{QP}|\psi_P>=\sum_i^N 
\{ <\Phi|G^+|\Phi>\}^{-1}_{ji}<i|\psi_0>.
\end{equation}
The  solutions of Eq.\,(\ref{eqn_III.19})  
can be written in terms of $|\psi_0>$ as 
\begin{equation}\label{eqn_III.22}
|\psi_P>\ =\ |\psi_0> -\sum_{ij}^N G^+|i>
\{ <\Phi|G^+|\Phi>\}^{-1}_{ij}<j|\psi_0>\
=\ |\psi_0>-\sum_{ij}^N G^+ \Lambda_{ij}|\psi_0>,
\end{equation}
wherein one can identify a separable potential
\begin{equation}\label{eqn_III.23}
|i>\{<\Phi|G^+|\Phi>\}^{-1}_{ij}<j|\ =\ |i>
\lambda_{ij}<j|=:\Lambda_{ij}(r,r')
\end{equation}
Note then that definition of  Q space gives  a 
 specification of the separable  
strengths $\lambda_{ij}(lsj,E)$ that is unique.
 The resultant Eq.\,(\ref{eqn_III.22}) has the 
form of a first order Born approximation but in 
fact it is an  exact result.

To proceed, we initially abandon the exactitude  of
Eq.\,(\ref{eqn_III.22}) and require the strength matrix, 
\begin{equation}\label{eqn_III.24}
\lambda_{ij}\ =\ \{<\Phi|G^+|\Phi>\}^{-1}_{ij},
\end{equation}
to be  constrained asymptotically by the experimental
 S-matrix of the full Hamiltonian
Schr\"odinger equation, {\em i.e.} asymptotically we 
induce  $|\psi_P>=|\Psi_{\cal H}>$.
This implies that  complex optical model
strengths $\lambda_{ij}$ emerge 
as a result of matching to Riccati-Hankel functions 
and non unitary S-matrices with 
\begin{equation}\label{eqn_III.25}
|\Psi_{\cal H}>= |\psi_P>\sim {1\over 2i}\left[-h^-(rk) 
+h^+(rk) S(k)\right].
\end{equation}
The strengths $\lambda_{ij}$ then can be simply determined from the 
linear system of equations 
\begin{equation}\label{eqn_III.26}
\frac 1{2i}{h^+(Rk)\left[S(k)-S_0(k)\right]= 
\sum_{ij}G^+|i>\lambda_{ij}<j|\psi_0^+>}.
\end{equation}
To reinforce a Lippmann--Schwinger equation, with 
the experimental S-matrix
as   boundary condition or equivalently with 
strengths $\lambda_{ij}$ from Eq.\,(\ref{eqn_III.26}), 
a transformation  of the separable potential 
Eq.\,(\ref{eqn_III.23}) is made.
This is achieved with 
\begin{equation} \label{eqn_III.27}
{\cal V}(r,r'):=\Lambda{1\over(1-G^+\Lambda)},
\end{equation}
which contains the separable potentials as defined with
Eq.\,(\ref{eqn_III.23}) but whose strengths 
now are solutions of Eq.\,(\ref{eqn_III.26}).  
As the transformation Eq.\,(\ref{eqn_III.27}) 
contains  integration of orthonormal functions,
 only strengths are altered. Using this optical 
model in  
the full Hamiltonian,
physical solutions are obtained with 
reference solutions $|\psi_0>$ and   Greens function $G^+$ of the
reference Hamiltonian $H_0$ by means of the 
Lippmann--Schwinger equation
\begin{equation}\label{eqn_III.28}
|\Psi_{\cal H}>=|\psi_0>+G^+{\cal V}|\Psi_{\cal H}>.
\end{equation}

\subsection{Technical details}

The partial wave radial wave functions 
of the reference potential satisfy  equations
\begin{equation} \label{eqn_III.29} 
u''_\alpha(r,k)=\left[{\ell(\ell+1)\over r^2}+{2\mu\over \hbar^2 } 
{V_a(r)\over 1+2V_b(r)}-\left({V_b'(r)\over 1+2V_b(r)}
\right)^2-{k^2\over  
1+2V_b(r)}\right]  u_\alpha(r,k), 
\end{equation}
wherein we identify the complete  set of
quantum numbers by the subscript $\alpha$. 
These equations  
we have solved numerically  for uncoupled and coupled channels
using a Numerov method. The
potentials $V_a,V_b,V_b'$ are  dependent on 
 the quantum numbers ($\ell,s,j$)
and are taken from the Paris, Nijmegen, Argonne, 
and  inversion  r-space potentials
as one wishes. The
Paris and Nijmegen-I are  momentum dependent 
potentials
with  $V_b\neq 0$, while
the  Nijmegen-II, Reid93, AV18, and inversion 
potentials all have $V_b=0$.
The physical solutions are matched
asymptotically to Riccati-Hankel functions
\begin{equation} \label{eqn_III.30}
u^+_\alpha(r,k)\sim \frac {1}{2i}\left[-h^-_\alpha(rk)+h^
+_\alpha(rk)S^0_\alpha(k)\right]
\end{equation}
and normalized by 
\begin{equation} \label{eqn_III.31}
\psi_\alpha^+(r,k)={u_\alpha^+(r,k)\over\sqrt{ 1+2V_b(r)}}.
\end{equation}
The irregular outgoing wave Jost solutions 
\begin{equation} \label{eqn_III.32}
{\cal J}^+_\alpha(r,k)\sim h^+_\alpha(rk)
\end{equation}
are calculated in the same way as the physical ones  and
they define the reference potential Green functions by 
\begin{equation} \label{eqn_III.33}
G^+_\alpha(r,r',k)=\left\{\begin{array}{l}
-\left(2\mu/ \hbar^2 \right)\displaystyle{1 \over k}
\psi^+_\alpha(r,k){\cal J}^{+ T}_\alpha(r',k),\quad  r<r' \\[0.3cm]
-\left(2\mu/ \hbar^2 \right)\displaystyle{1 \over k}
{\cal J}^+_\alpha(r,k){\psi}^{+ T}_\alpha(r',k),\quad  r>r' ,
\end{array}\right.
\end{equation}
where the transpose matrix is signaled by the superscript $T$. 
At the asymptotic matching radius $R$
\begin{equation} \label{eqn_III.34}  
\Psi_\alpha^+(R,k)=\psi^+_\alpha(R,k)+
\int_0^\infty G^+_\alpha(R,r_1,k)\Phi_\alpha(r_1)
dr_1\lambda_\alpha(k)
\int_0^\infty \Phi_\alpha(r_2)\psi^+_\alpha(r_2,k)dr_2 ,
\end{equation}
and taking the difference between the reference 
and full S-matrix, this reduces to 
\begin{eqnarray} \label{eqn_III.35} 
\Psi_\alpha^+(R,k)-\psi^+_\alpha(R,k)&=&
\frac 1{2i}h_\alpha^+(Rk)\left[S_\alpha
(k)-S^0_\alpha(k)\right]\nonumber \\[0.2cm] 
&=&
\int_0^\infty G^+_\alpha(R,r_1,k)\Phi_\alpha(r_1)dr_1 
\lambda_\alpha(k)
\int_0^\infty\Phi_\alpha(r_2)\psi^+_\alpha(r_2,k)dr_2 .
\end{eqnarray}
A linear expression for the potential
strength $\lambda_\alpha(k)$ results.  The strengths  
are transformed by 
Eq.\,(\ref{eqn_III.27})
to  give final separable potential strengths
\begin{equation}\label{eqn_III.36}
\sigma_\alpha(k)=\left(1-\lambda_\alpha(k)
\int_0^\infty\int_0^\infty\Phi_\alpha(r_1)G^+_\alpha(r_1,r_2,k)
\Phi_\alpha(r_2)\,dr_1\, dr_2
\right)^{-1}\lambda_\alpha(k).
\end{equation}
These strengths $\sigma_\alpha(k)$  define the proper optical 
model of Eq.\,(\ref{eqn_III.28}),
given for the more general coupled channel and rank $\leq3$ 
separable potentials, to be
\begin{equation} \label{eqn_III.37}
{\cal V}(r,r')=|\Phi_\alpha>\sigma_\alpha<\Phi_\alpha|=
{ 2\mu \over \hbar^2 }\Phi_\alpha {\cal W_\alpha}\Phi_\alpha^T
\end{equation}
where
\begin{equation} \label{eqn_III.38} 
\Phi_\alpha:=
\left\{ \begin{array}{cccccc}
\Phi^1_{j-1/2}(r)&\Phi^2_{j-1/2}(r)&\Phi^3_{j-1/2}(r)& 0 & 0 & 0\\
 0 & 0 & 0 & \Phi^1_{j+1/2}(r)&\Phi^2_{j+1/2}(r)&\Phi^3_{j+1/2}(r)
\end{array}\right\},
\end{equation} 
and the  symmetric strength matrices are 
\begin{equation} \label{eqn_III.39}
{\cal W}_\alpha(k):=\mbox{Re} W_{i,j}+i\mbox{Im} W_{i,j}=
(\hbar^2/2\mu)\ \sigma_\alpha(k),
\mbox{\ for\ }i,j=1\cdots 6.
\end{equation}  
For single channel and rank one potentials, 
this representation is obviously reduced.

There are several options one may consider for the 
separable potential form factors
$\Phi_\alpha(r)$. First,  any finite rank potential may be
chosen with the strengths $\lambda_\alpha(k)$ determined from data
at several energies around a mean energy. In practice, 
using a rank $>1$ 
option has been successful for single channels  
but inherent  lack of
energy dependence for coupled channels strongly favors restricting 
potentials to be of rank one. Next is the choice of radial
form factors. As rank one potential form factors 
we have used a.) normalized 
harmonic oscillator radial wave functions  
$\Phi_\alpha=\Phi_\ell (r,\hbar\omega)$,
b.) normalized Gaussian functions 
$\Phi_\alpha=N_0\exp -(r-r_0)^2/a_0^2$ with
$r_0$ and $a_0$ being parameters, c.) a normalized edge  
function 
$\Phi_\alpha (r_0)=1/2h, \Phi_\alpha(r_0\pm h)=1/4h$ and 
$\Phi_\alpha(r,\alpha)=0$ otherwise, and 
d.) a boundary condition model  realized  by  $
\Phi_\alpha(r_0)=1/h$ and 
$\Phi_\alpha(r)=0$ otherwise.
The last option is suitable for a sudden transition 
from the hadronic domain
into the QCD domain and back. Of course these are 
but examples and 
others may be inspired by more explicit considerations of QCD.

Solutions of the full problem Lippmann--Schwinger 
equation, Eq.\,(\ref{eqn_III.28}), 
have been  generated with reference potential solutions 
 and  Green functions 
as per Eq.\,({\ref{eqn_III.33}) and with  separable
potentials  whose strengths $\sigma_\alpha(k)$ are 
given by Eq.\,(\ref{eqn_III.36}).
These solutions are readily found from systems of linear
equations, for single and coupled channels, using a
trapezoidal integration rule for Eq.\,(\ref{eqn_III.28}) 
recast as
\begin{equation}\label{eqn_III.40}
\Psi_\alpha^+(r)=\psi^+_\alpha(r) +\int_0^\infty G^+_\alpha
(r,r_1)\Phi_\alpha(r_1)dr_1{\cal W}_\alpha(k) \;
\int_0^\infty \Phi_\alpha(r_2)\Psi^+_\alpha(r_2)\;dr_2 \ .  
\end{equation}
However there is a faster method by which solutions 
Eq.\,(\ref{eqn_III.34}) as well as half off-shell
wave function solutions and t-matrices can be found. 
This we consider next.

\subsection{Evaluation of the half off-shell t-matrix}
 
The calculation of NA optical potentials for 
$T_{Lab} < 3$ GeV requires 
half off-shell  NN t-matrices for  on-shell k-values  $k<6$ fm$^{-1}$,
and a correspondingly large range of off-shell values. 
In principle, 
in applications the off-shell k-values (later identified 
with $q$) are needed in 
integrals from $0 \to \infty$  but a reasonable upper limit is $q=2k$. 
A fast and stable method  of evaluation of such 
half off-shell t-matrices, when  r-space potentials are chosen, is 
an extension of the Schr\"odinger equation as an inhomogeneous
differential equation. The method follows that of
Van Leeuwen and Reiner \cite{vLR61}.
 
The most general potentials in our study  contain  
momentum dependent, local and 
separable complex potentials  for both  single and coupled channels. 
In particular for the results shown,  a rank one separable potential 
with a radial harmonic oscillator form factor 
$\Phi_\ell (r,\hbar\omega)$ has been used.
The Schr\"odinger equation then can be cast as 
\begin{eqnarray}\label{eqn_III.41}
\left[ {\cal M}(r)\frac{d^2}{dr^2}-{\cal M}(r){\ell(\ell + 1)\over r^2} 
- V_a(r) + V_b''(r) + 2 V_b'(r) \frac{d}{dr}+k^2 \right]
\psi_\ell (r,k,q)\nonumber\\[0.2cm]
=\Phi_\ell (r)\lambda_\ell (k^2)\int_0^\infty 
\Phi_\ell (x)\psi_\ell (x,k,q)\,dx
+(k^2 - q^2)j_\ell (rq),\qquad 
\end{eqnarray}
with ${\cal M}(r)=(1+2\,V_b(r))$.
The regular solutions of which not only must 
vanish at the origin
but also asymptotically  must match to
\begin{equation} \label{eqn_III.42}
\lim_{r\to\infty} \psi^{(\pm,0)}_\ell (r,k,q){\cal N}_\ell  = 
j_\ell (rq)+h_\ell ^{(\pm,0)}(rk){q\over k}T_\ell ^{(\pm,0)}(k^2,k,q)
\end{equation}
to determine the half off-shell t-matrix 
$T_\ell ^{(\pm,0)}(k^2,k,q)$
and the normalization ${\cal N}_\ell $. 
Spherical Riccati functions are symbolized by $j_\ell (x)$,  
$h^\pm_\ell (x)$ and $h^0_\ell(x)=n_\ell (x)$. 
In the following we suppress the
channel subscript $\ell$ as the expressions hold 
for single and coupled channels. 
The on-shell t-matrix gives the S-matrix by the relation 
\begin{equation}\label{eqn_III.43}
S(k) = 1 + 2 i\,  T^{(+)}(k^2,k,k).
\end{equation}
To solve for coupled channels $^3SD_1,\ ^3PF_2,$ {\em etc.},   
two linear 
independent regular solutions are calculated and 
Eqs.\,(\ref{eqn_III.36}), (\ref{eqn_III.41}) and 
(\ref{eqn_III.42}) are to be understood as  $2\times 2$ 
matrix equations.

The regular solutions are readily found numerically as follows.
First, a regular solution of the reference potential 
Schr\"odinger equation 
\begin{equation} \label{eqn_III.44}
f_0''(r,k) - V(r)f_0(r,k)+k^2f_0(r,k)=0
\end{equation}
is calculated. Therein $V(r)$ implies all the local 
potential terms including the
centripetal barrier. Then a regular solution of the
full potential Schr\"odinger equation with the 
reference potential $V(r)$ 
and separable potential,  
\begin{equation} \label{eqn_III.45}
f_1''(r,k)-V(r)f_1(r,k)+k^2f_1(r,k)=\Phi(r)
\sigma(k)<\Phi|f_1>,
\end{equation}
is obtained from  a particular solution of
\begin{equation} \label{eqn_III.46}
g_1''(r,k)-V(r)g_1(r,k)+k^2g_1(r,k)=\Phi(r)
\sigma(k){\cal F},
\end{equation}
where we use ${\cal F}=<\Phi|f_0>$, and 
$f_1(r,k)=f_0(r,k){\cal A}+g_1(r,k)$. The factor 
(matrix) $\cal A$
is determined from  
\begin{equation} \label{eqn_III.47}
<\Phi|f_1>=<\Phi|f_0>{\cal A}+<\Phi|g_1>
\end{equation}
and 
\begin{equation} \label{eqn_III.48}
{\cal A}=1-{\cal F}^{-1}<\Phi|g_1>.
\end{equation}
Finally the regular solution $f_1(r,k)$ 
can be multiplied with any complex number 
(matrix) to be 
a general regular  solution of Eq.\,(\ref{eqn_III.45}).

The half off-shell t-matrix is related to  the regular half 
off-shell wave function $\psi(r,k,q)$, which satisfies 
the inhomogeneous Schr\"odinger equation 
\begin{equation} \label{eqn_III.49}
\left[{d^2\over dr^2}-V(r)+k^2\right]\psi(r,k,q)=
\Phi(r)\sigma(k)<\Phi|\psi>+(k^2-q^2)j(rq).
\end{equation}
Asymptotically  this wave function is 
\begin{equation} \label{eqn_III.50}
\psi^{(\pm,0)}(r,k,q)\sim j(rq)+h^{(\pm,0)}(rk)
{q\over k}T^{(\pm,0)}(k^2,k,q).
\end{equation}
A general regular solution of
Eq.\,(\ref{eqn_III.45}) and a particular regular  inhomogeneous 
solution of Eq.\,(\ref{eqn_III.43})
then is  needed to satisfy the boundary conditions
 given in Eq.\,(\ref{eqn_III.50}). 
A particular solution of Eq.\,(\ref{eqn_III.49}) is  obtained 
in two steps. First, with
\begin{equation} \label{eqn_III.51}
{\cal F}=<\Phi|g_2>=<\Phi|f_1>=<\Phi|f_0>,
\end{equation}
a particular solution is given by
\begin{equation} \label{eqn_III.52}
f_2(r,k,q)=f_1(r,k){\cal B}+g_2(r,k,q),
\end{equation}
where $\cal B$ is determined from
\begin{equation} \label{eqn_III.53}
{\cal B}=1-{\cal F}^{-1}<\Phi|g_2>.
\end{equation} 
The off-shell wave function matches asymptotically as 
\begin{equation} \label{eqn_III.54}
\psi^{(\pm,0)}(r,k,q)=f_1(r,k){\cal N}+f_2(r,k,q)\sim j(rq)+
h^{(\pm,0)}(rk){q\over k}T(k^2,k,q).
\end{equation}
The normalization  $\cal N$ and t-matrix $T^{(\pm,0)}(k^2,k,q)$ are 
readily evaluated from the {\em quasi} Wronskians 
\begin{eqnarray} \label{eqn_III.55}
{\cal N}&=&W^{-1}[h^{(\pm,0)},f_1]
\left(W[j,h^{(\pm,0)}]-W[h^{(\pm,0)},f_2]\right)
\nonumber\\[0.2cm]
{q\over k}T^{(\pm,0)}(k^2,k,q)&=&
W^{-1}[j,h^{(\pm,0)}]\Bigl( W[j,f_1]{\cal N}+W[j,f_2]\Bigr).
\end{eqnarray}
where we define 
\begin{equation} \label{eqn_III.56}
W[a,b]:={(a_n-a_{n-1})\over h}b_n-a_n{(b_n-b_{n-1})\over h}
\end{equation}
at two asymptotic radial points $r_{n-1}$ and $r_n=r_{n-1}+h$. 
The quantities $a$ and $b$ can be either scalars or  matrices.

It is very convenient to use the Numerov algorithm
to solve Eqs.\,(\ref{eqn_III.44}), 
(\ref{eqn_III.46}), and (\ref{eqn_III.49}). But to do so
for Eq.\,(\ref{eqn_III.41}) requires 
equations without first derivative terms. 
The above can be made so by use of a factorization  
\begin{equation}\label{eqn_III.57}
\psi(r,k,q)=f(r,k,q){\cal D}(r),\mbox{\ with\ \ } {\cal D}(r)=
{1\over \sqrt{1+2V_b(r)}}
\end{equation} 
The resulting equation for $f(r,k,q)$ is
\begin{eqnarray}\label{eqn_III.58} 
f''(r,k,q)=\left[ {\ell(\ell+1)/r^2}-{\cal D}(r)k^2{\cal D}(r)+
{\cal D}(r)V_a(r) {\cal D}(r)+
\Biggl( {\cal D}(r)V_b'(r){\cal D}(r) \Biggr)^2\right]f(r,k,q)
\nonumber\\[0.2cm]
+\Phi(r) {\cal D}(r)\sigma(k)<{\cal D} \Phi|f>
+ (k^2-q^2)j_\ell (rq) {\cal D}(r).\qquad
\end{eqnarray}

\subsection{Numerov Algorithm}

The solution of radial Schr\"odinger equations is certainly 
not new and generally deserves
no mention. Here, we dwell upon the details since we found 
the specified elements
to have a {\em normal form} of related problems in other fields of
physics and engineering which were tested with parallel computing 
facilities.
The Numerov algorithm has been  widely used for single and coupled channels
Schr\"odinger equations since it gives sufficient  numerical 
accuracy with  
minimal  operations \cite{Ray00}. The standard form of linear 
homogeneous or 
inhomogeneous Schr\"odinger equations which we have to solve is
\begin{equation} \label{eqn_III.59}
f_i''(r)=V_{ij}(r)f_j(r)+W_i(r),
\end{equation}
where $W_i(r)=0$ for homogeneous equations.
The terms  $V_{ij}(r)$ and $W_i(r)$ are easily identified in 
Eq.\,(\ref{eqn_III.58}).
For  single channels the algorithm is 
\begin{equation}\label{eqn_III.60}
f_{n+1}=2f_{n}-f_{n-1}+{h^2\over 12}\left( u_{n+1}+10 u_{n}+u_{n-1}\right)
\end{equation}
or 
\begin{eqnarray}\label{eqn_III.61}
\left( 1- {h^2\over 12}V_{n+1}\right) f_{n+1}&=&
\left( 2+ {10 h^2\over 12}V_{n}\right) f_{n}
-\left( 1- {h^2\over 12}V_{n-1}\right) f_{n-1}
\nonumber \\[0.2cm]
&+&  {h^2\over 12}\left(W_{n+1}+10\,W_{n}+W_{n-1}\right)\ .
\end{eqnarray}
These expressions generalize for coupled channels 
using  standard  
vector and matrix algebra. A significant reduction 
of operations is found
by using the substitution 
\begin{equation}\label{eqn_III.62}
\xi_n=\left(1-{h^2\over 12}V_n\right)f_n 
\end{equation}
in Eq.\,(\ref{eqn_III.61}). It  gives
\begin{equation}  \label{eqn_III.63}
\xi_{n+1}=2\xi_n-\xi_{n-1}+{\cal U}_n,
\end{equation}
and the inhomogeneous equation 
\begin{equation} \label{eqn_III.64}
\xi_{n+1}=2\xi_n-\xi_{n-1}+{\cal U}_n +
{h^2\over 12}\left(W_{n+1}+10W_n+W_{n-1}\right) ,
\end{equation}
with 
\begin{equation} \label{eqn_III.65}
{\cal U}_n={h^2 V_n\over 1-{h^2\over 12}V_n}\xi_n.
\end{equation}
Back-transformations from $\xi_i \to f_i$ use either of the
two possibilities 
\begin{equation}\label{eqn_III.66}
f_i=\xi_i+\frac1 {12}{\cal U}_i,\mbox{\ or}\quad 
f_i={\xi_{i+1}+10\xi_i+\xi_{i-1}\over 12}.
\end{equation}


\section{Properties and discussion of the optical model.}

A range of optical potentials have  been generated using the  
algorithm developed above.
As reference potentials, the  Paris, Nijmegen, Argonne, and 
inversion potentials
have been used.
For the separable potential form factors,
normalized harmonic oscillator functions (HO), 
$\Phi_\ell (r,\hbar\omega)$, with  $200 <\hbar\omega < 900$ MeV 
have been used.
The same $\hbar\omega$ is used for all partial waves however.
For single channels all quantum sets with   $J\leq 7$ were 
included while
those for   $J\leq 6$ were used with the coupled channels. 
A superposition
of several HO functions  with radial quantum numbers $n=1,2,3$ 
was allowed and with data in intervals $T_{Lab}\pm 25$, $T_{Lab}\pm50$, 
and $T_{Lab}\pm 100$ MeV,  optimal solutions of 
Eq.\,(\ref{eqn_III.35})
found using a least square linear equation routine 
from a scientific subroutine
library (NAG-library). 
This procedure was used to
determine a single $\hbar\omega$ for energies within 
$0.5<T_{Lab}<2$ GeV with an 
overall low $\chi^2$. That  optimal 
value is $\hbar\omega=450$ MeV. For higher energies 
$2<T_{Lab}<3$ GeV
and low partial waves, this optimal oscillator has 
bound states embedded in the continuum, but as such they
are of no concern in this analysis and so 
we used  $\hbar\omega=450$ MeV for all energies $0.3<T_{Lab}<3$ GeV. 
With rank one separable potentials, the HO  functions 
(radial quantum number $n=1$) are 
\begin{equation}
\Phi_\ell (r,\hbar\omega)\sim r^{\ell+1}\exp -(r/r_0)^2,
\mbox{\ with\ \ }r_0=\sqrt{{2\hbar^2\over \mu \hbar\omega}},
\end{equation}
and with $\hbar\omega=450$ MeV,  $r_0=0.61$ fm.
Then with the separable form fixed, it is trivial to solve  
Eq.\,(\ref{eqn_III.35}) with S-matrix data 
taken at each energy.
\begin{figure}[tbp]
\centering
\epsfig{file=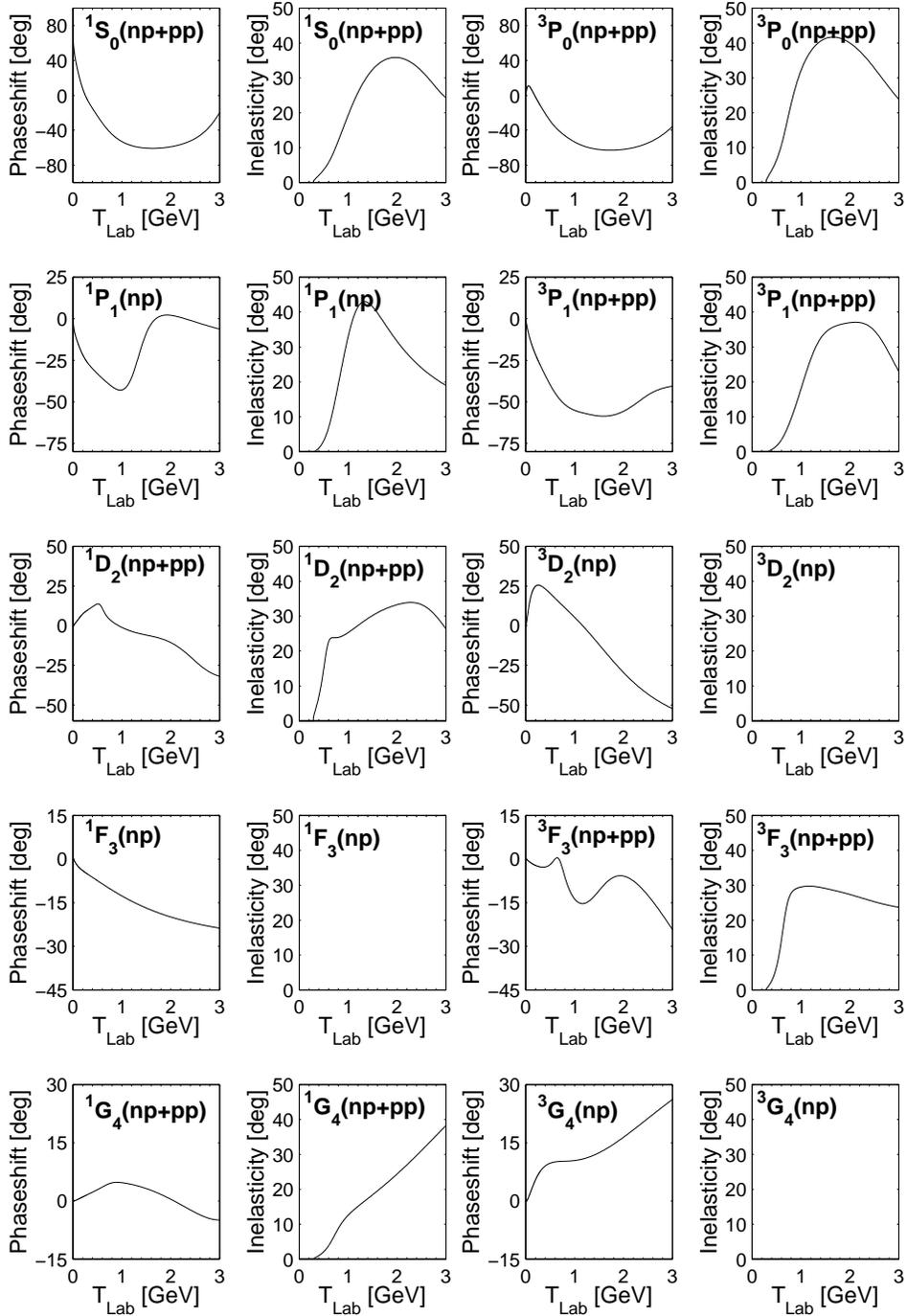,scale=0.75}
\caption{SP00 phase shifts for  $np$  single channels.}
\label{figure_IV.11}
\end{figure}
In Figs.\,\ref{figure_IV.11} and 
\ref{figure_IV.12}
\begin{figure}[tbp]
\centering
\epsfig{file=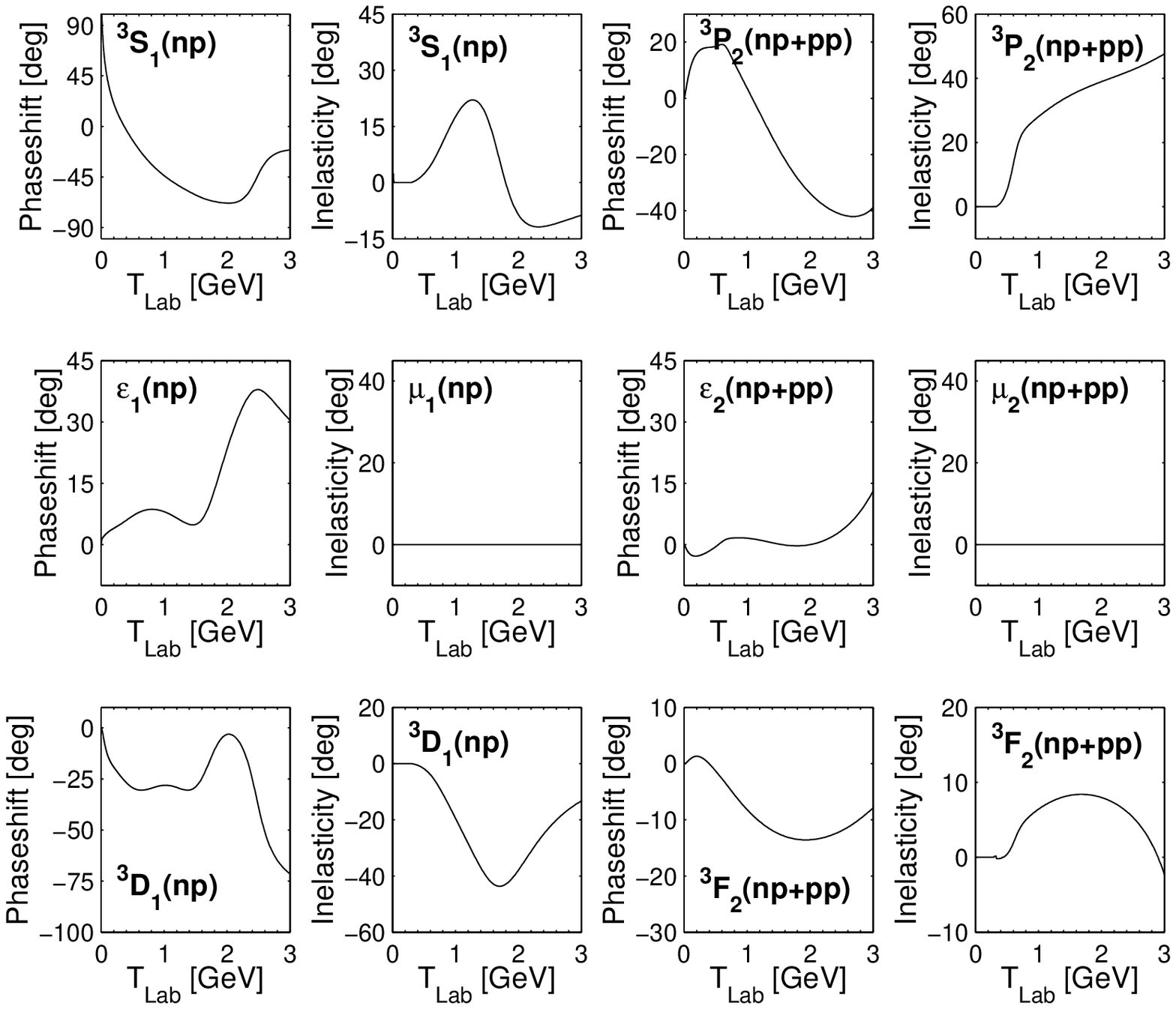,scale=0.75}
\caption{SP00 phase shifts for  $np$  coupled channels.}
\label{figure_IV.12}
\end{figure}
we show the full potential
model phase shifts that result on solving scattering from 
the deduced optical potentials. 
They are {\em identical } with the SP00 solution. 

\begin{figure}[tbp]
\centering
\epsfig{file=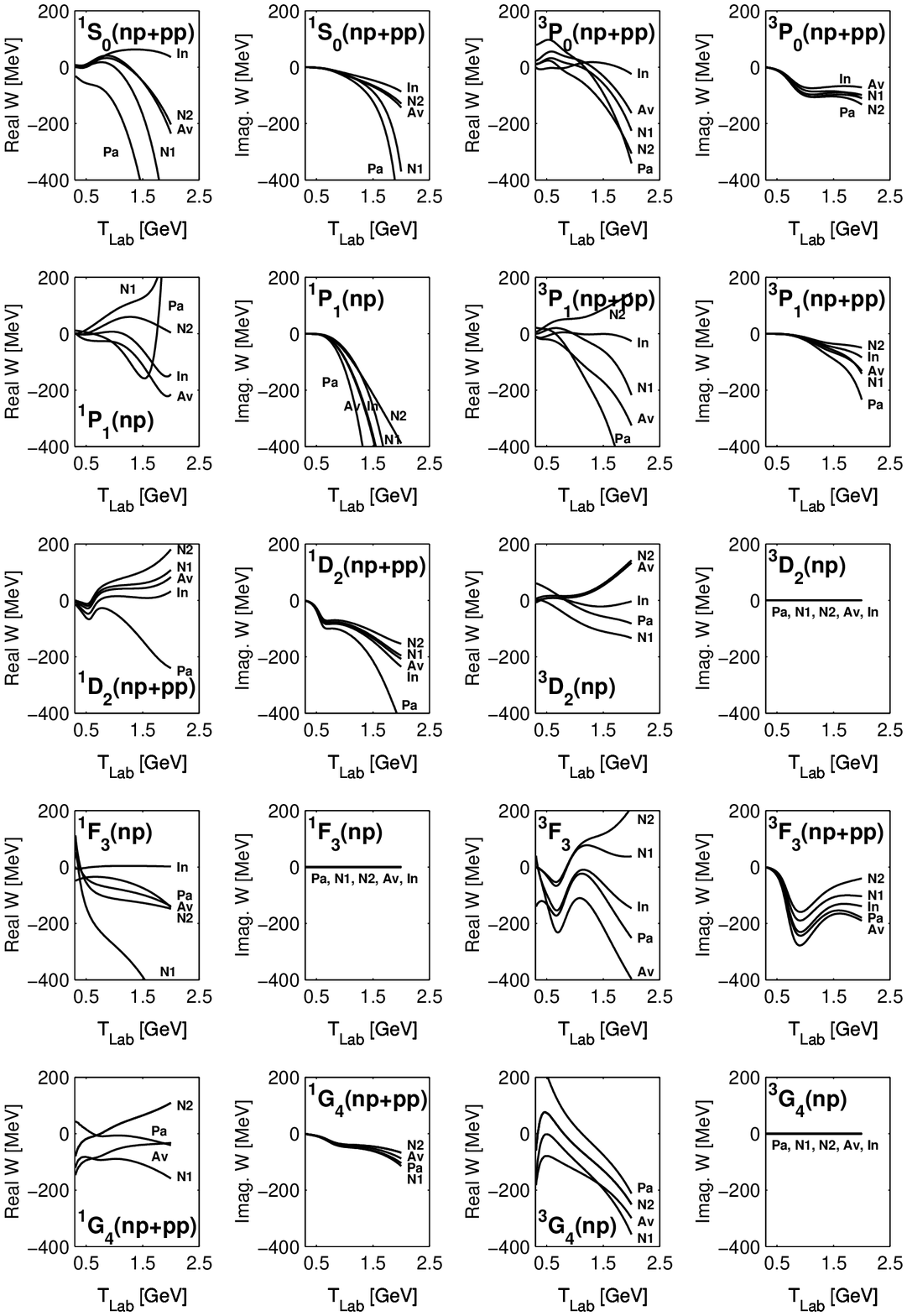,scale=0.7}
\caption{$np$  single channel separable potential strengths, using  
inversion (In), Paris (Pa), Nijmegen (N1, N2) and Argonne AV18 (Av)  
as reference potentials with  
$\hbar\omega=450$ MeV.}
\label{figure_IV.13}
\end{figure}
The strengths $\sigma_\alpha(k)$ of Eq.\,(\ref{eqn_III.36}) 
were determined independently 
for each given reference potential
and the optical potential values of Eq.\,(\ref{eqn_III.39})
are shown in Figs.\,\ref{figure_IV.13} and \ref{figure_IV.14}.
\begin{figure}[tbp]
\centering
\epsfig{file=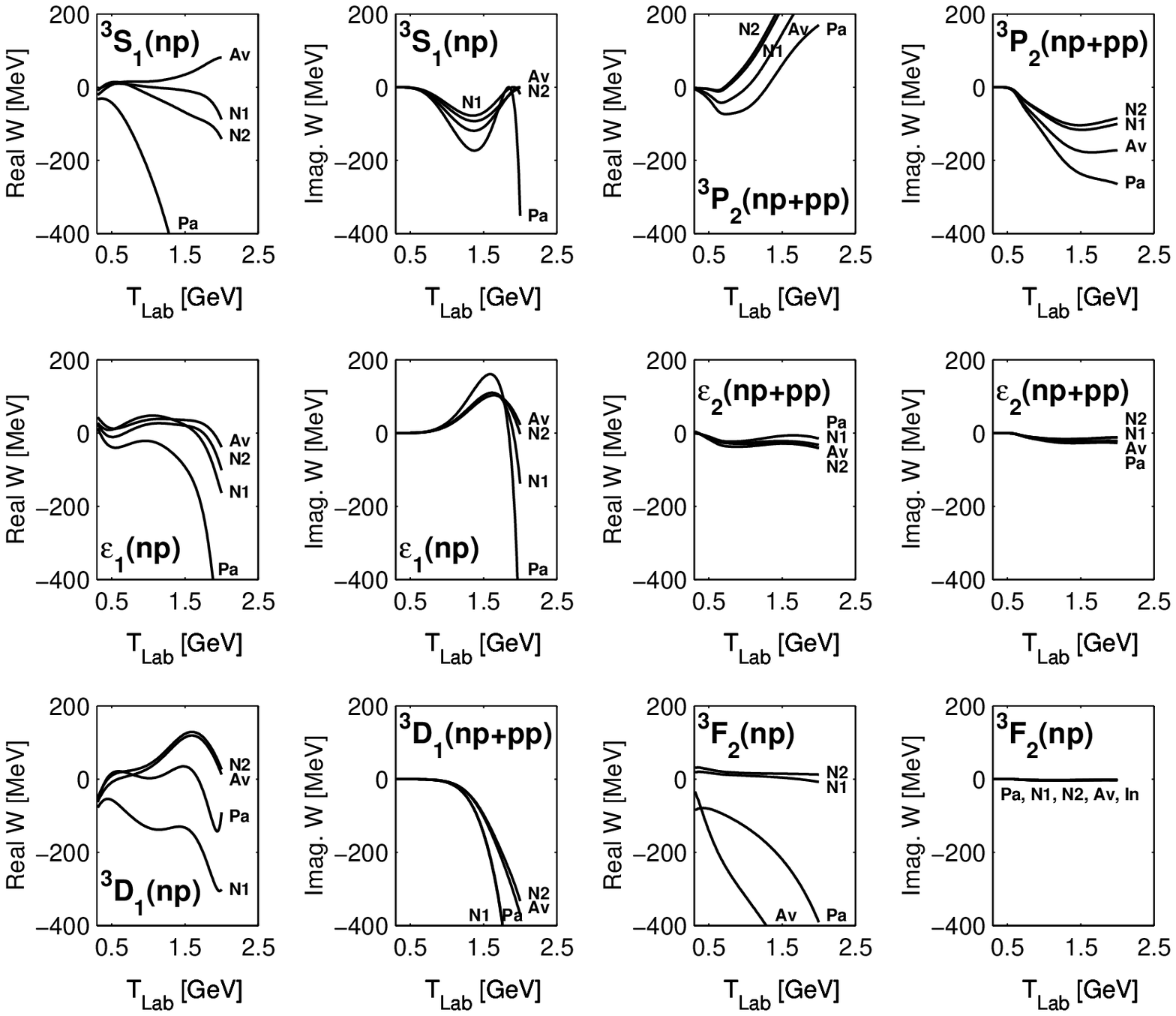,scale=0.7}
\caption{$np$  coupled channel separable potential strengths, using  
Paris (Pa), Nijmegen (N1, N2) and Argonne AV18 (Av)  
as reference potentials with   
$\hbar\omega=450$ MeV.}
\label{figure_IV.14}
\end{figure}
The lettering in the small sub-figures
identify the channel.  The curves give the results obtained 
when the 
Paris (Pa), Nijmegen-I (N1), Nijmegen-II (N2), Argonne AV18 
(Av) and single channel
inversion potentials (In) were used as reference potentials.
These optical model strengths display two most important features. 
The first is
that they are not insignificant. The reference potentials by
 themselves fail 
to account for the phase shifts $\delta$ and 
$\delta^\pm,\varepsilon$. The second feature of importance is the 
loss of unitarity of the S-matrices
accounted for by $\rho$ and $\rho^\pm,\mu$. The two features are weakly
coupled by the optical and reference potentials respectively. 
Below threshold however, a purely  real optical potential and
 very small strengths 
reflect the agreement of the reference potential phase shifts  with SP00. 
The imaginary potentials show a  smooth
energy dependence starting at threshold  $T_{Lab}=280$ MeV and,
 by  having negative
values,  account for flux loss. Notice also that the results 
using inversion reference
potentials (In) in the channels 
$^1S_0,\ ^3P_0,\ ^3P_1,\ ^1P_1,\ ^3D_2$ and $^1F_3$ have small
values for the optical potential real strengths. 
Thus those real potentials need  hardly any  modifications 
at short distances and this
supports the conjecture of the soft core potential 
discussed in Sect.\,II regarding Fig.\,\ref{figure_II.7}. 
All reference potentials are most uncertain in the 
$^1D_2$, $^3F_3$ and $^3PF_2$ channels. 
This is well known as the region  300 MeV to 1 GeV is 
dominated by the $\Delta$(1232) 
resonance while many $N^*$ and higher spin resonances 
shape the region  1 to  2 GeV. 
Indeed the  obvious energy dependences seen in the $^1D_2$ and $^3F_3$
channels are signatures of the strong coupling to 
the $\Delta$(1232) resonance between 
$T_{Lab}=500$ and  $750$ MeV. The coupled channel results shown in 
Fig.\,\ref{figure_IV.14} follow closely the conclusions 
drawn for the single channel results. Thus only the 
$^1D_2$ and $^3F_3$ channels
show energy dependences in the real phase shifts 
$\delta$ and absorptions $\rho$ that 
require particular attention and an  explicit treatment 
of resonance coupling. The $^3PF_2$
coupled channels show some similar $\Delta$(1232) 
resonance coupling around 600 MeV. 
All the other channels support an energy independent 
local  reference potential which can 
be generated by Gel'fand--Levitan--Marchenko inversion
using the real phase shift data. Also, as  the optical 
potential strengths vary
smoothly with energy for these channels,  use of  a complex but local 
very  smoothly energy dependent complex
potential with Gaussian or Yukawa form factors is suggested \cite{Ger98}. 
It may be that within  QCD hybrid models such a 
{\em local background optical potential}
can be formulated microscopically and be linked
with the high energy diffraction and Regge models of elastic
scattering \cite{Jen00,Lo87,Mat94,Col77}. 

In addition to those optical model potentials 
found by using $\hbar\omega=450$ MeV, calculations
where also made using $\hbar\omega=750$ and $900$ MeV. 
This increase in $\hbar\omega$ reduces
the range $r_0$ from $0.61 \to 0.47$ and $\to 0.43$ fm
respectively. 
The primary purpose of those calculations was a search 
of the effective radial domain 
in which the reference potentials all differ most markedly. 
A shorter range of the form factor $\Phi_\alpha$ leads to 
increased values of the optical potential
strengths and thus the shortcomings of the reference potentials are
magnified. To interpret this magnification, one must bear in mind  
the boundary condition on wave functions
to be zero at the origin  and the influence of a potential in the short
range region, between $0<r<0.8$ fm for the results 
found using $\hbar\omega=450$ MeV
and $0<r<0.5$ fm in the case of $\hbar\omega=750$ MeV. 
Of note in these calculations 
is that only on  using the inversion potentials as reference do
the real optical model strengths   remain small.
Given that the inversion potentials were designed by 
themselves to give the SP00 real
phase shifts
up to $T_{Lab}=3$ GeV as derived from the real parts 
of the K-matrix Eq.\,(\ref{eqn_II.1}), 
that aspect lends further support for a decoupling of 
the real and imaginary parts of the optical model
potentials in calculations. 
Interference effects are small with the implication  
that the real and imaginary
parts might be independently assessed. Such is not so 
evident when the OBEP are used as the
reference potentials and the particular poor 
extrapolations one finds on using the
Paris and Nijmegen-I that have explicit momentum  
dependences are most noticeable.

The  $^1D_2$ and $^3F_3$ channel results are exceptional. 
Even with the
inversion potentials as reference, the modulating optical 
potentials have
comparable real and imaginary parts. Such reflect the means by which the 
optical model accounts for specific strong resonance effects. 

The changes wrought in complex potential correction 
strengths when any OBEP
is used as reference and  when the $\hbar\omega$ for the defining 
optical potential correction 
form factors is enlarged to 750 and 900 MeV respectively, 
further stresses that any are
poor choices as reference as one forces their 
{\em propriety} to even shorter
ranges. In sum such have scant credibility in the range $0.5<r<1$ fm.  
 
Our studies support the conceptualization of the formation and
fusion  of two nucleons, more generally of two elementary particles 
like $\pi N$, $\pi\pi$,
{\em etc.}, into a combined object \cite{Ger00}. Such is correlated with
selective  enhancements of 
{\em probability density} and with {\em loss of flux} 
from the elastic scattering channel.
The probability density of the full  problem is 
\begin{equation} \label{eqn_III.67}
\rho_\alpha(r,k)  =  {1\over r^2}\ Trace\ \Psi^\dagger_\alpha (r,k)
\Psi_\alpha(r,k)
\end{equation}
and the  flux loss function, which  results  from the continuity
equation $\partial_t\rho_\alpha ({\bf r})+
({\bf \nabla\cdot j})_\alpha=0$ and the
time dependent Schr\"odinger equation, is
\begin{equation} \label{eqn_III.68}
({\bf\nabla\cdot j})_\alpha ={i\over \hbar} {1\over r^2} Trace
\int_0^\infty \{ \Psi^\dagger_\alpha(r,k)
{\cal V}_\alpha(r,r_1)\Psi_\alpha(r_1,k)
- \Psi^\dagger_\alpha(r_1,k)
{\cal V}^\dagger_\alpha(r_1,r)\Psi_\alpha(r,k)\}dr_1.
\end{equation}

\begin{figure}[tbp]
\centering
\epsfig{file=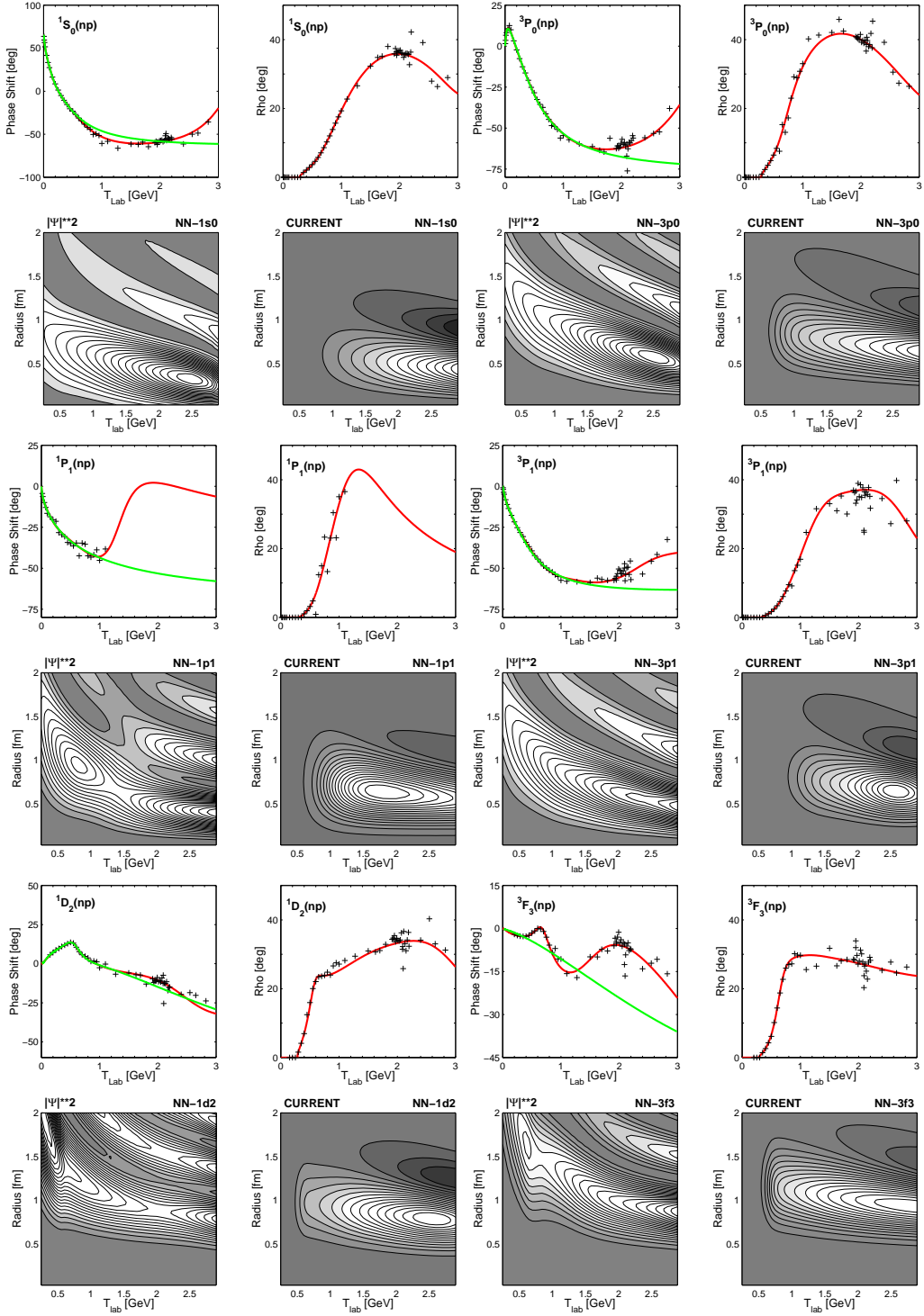,scale=0.8}
\caption{Block matrices containing  $\delta(T_{Lab})$ and $\rho(T_{Lab})$
of SP00, and energy versus radius the probability and  flux loss.}
\label{figure_IV.15}
\end{figure}
For several low partial waves, in Fig.\,\ref{figure_IV.15} 
we show probabilities as defined by Eq.\,(\ref{eqn_III.67}) 
and  flux loss via  Eq.\,(\ref{eqn_III.68}). In this figure the 
SP00 phase shift functions $\delta(T_{Lab})$ and 
$\rho(T_{Lab})$ are given as 
well for each channel and they are compared with the scatter of 
single energy solutions of SP00.  The inversion
potential phase shifts are given as well.
We show these single energy solutions to acknowledge 
their scatter about the
smooth SP00 solutions. Those sharp variations have 
been considered \cite{Bys87}
as evidence of narrow dibaryon resonances. 
Should they be so, we contend that our
potential model and associated viewpoint of fusion is still appropriate 
on geometric grounds. Such dibaryonic resonance effects require a detailed
QCD description of their structure and decay.

The contour plots give the probability distributions 
and the zonal flux losses for 
$0<r<2$ fm and $T_{Lab}<3$ GeV. From these contour 
plots we envisage 
a smooth development with energy for scattering in all channels
with possible exception of the $^1P_1$, $^1D_2$ and $^3F_3$ channels. 
Of those, the $^1P_1$ channel is bound by data only to 1.2 GeV,  
above this value the SP00 phase shift function is conjecture. 
Nevertheless we have used the solution to demonstrate what implication 
such a drastic variation of $\delta(T_{Lab})$ for 
$1<T_{Lab}<2$ GeV causes
in the probability distribution leaving the flux 
loss essentially  invariant.

The $^1S_0$ and $^3P_0$ results are given in 
Fig.\,\ref{figure_IV.15}. They have
very similar characteristics. The SP00 continuous energy solutions have 
phase shifts whose real parts have a minimum at about 1.6 GeV. 
The probability and flux loss plots show characteristic strongly
distorted structures with the short distance $0.25 <r<0.6$ fm  attributes
indicative of a large width ($\Gamma < 1$ GeV) resonance with strong
absorption. The $^3P_1$ results given in the bottom 
half of Fig.\,\ref{figure_IV.15}
are interpreted similarly. The $^1P_1$ results 
shown in this figure have more 
variation as the  resonance impact in the SP00 
solution is reflected in the 
flux loss plot in particular. The $^1D_2$ and 
$^3F_3$ channel results are given in 
Fig.\,\ref{figure_IV.15}. Concomitant
with the  structured SP00 phase shift functions the probability plots
indicate a change from the characteristic smoothness  
of the other channels with notable
features for $400<T_{Lab}<900$ MeV. A very long ranged 
probability peak with strong
distortions and  significant absorption extending 
beyond 1 fm is evident.

The details shown are not independent of the chosen  
geometry of the optical
potential but the patterns are quite stable with 
variations of the HO energy.  
These results support our pictorial conjectures of 
reaction schemes given in 
Figs.\,\ref{figure_II.5}, \ref{figure_II.9}  and
\ref{figure_II.10}. The energy dependence indicates 
that in the energy regime 
300 MeV to 1 GeV the
concept that one or the other of the colliding hadrons 
at most is excited to form the
$\Delta$-resonance  while the two hadrons remain 
as disparate entities. At higher 
energies, and for smaller radii, the strong absorption is consistent with a
fusion of the colliding particles.

The Kowalski--Noyes f-ratios of the half off-shell t-matrices
\begin{equation}\label{eqn_IV.4}
f_\alpha(k,q)={T^{(\pm,0)}_\alpha (T_{Lab}(k),k,q)
\over T^{(\pm,0)}_\alpha (T_{Lab}(k),k,k)}
\end{equation}
are useful quantities as they stress the  potential
 differences in momentum space.
For a purely real potential the Kowalski--Noyes
 f-ratio is real but this is no longer
the case for complex potentials. Nevertheless, 
the f-ratios are always  independent of the  boundary
conditions used in Eq.\,(\ref{eqn_III.42}) to
 determine $T_\alpha^{(\pm,0)}(k^2,k,q)$.
\begin{figure}[tbp]
\centering
\epsfig{file=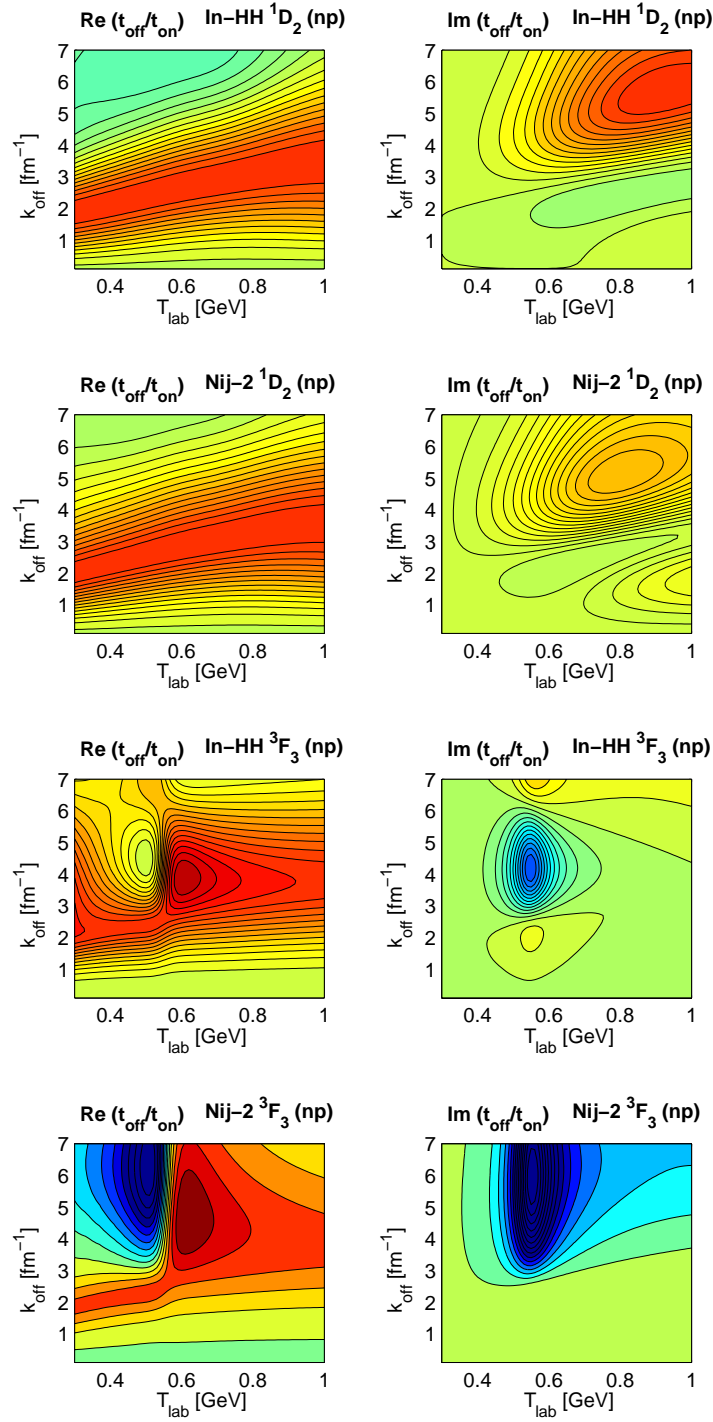,scale=0.76}
\caption{Kowalski-Noyes f-ratios, real part in left column and imaginary
part in right column, for the $^1D_2$ and $^3F_3$ channels 
calculated with inversion-HH and Nijmegen-II reference potentials and 
optical potentials using HO, $\hbar\omega=450$ MeV, separable form factor.}
\label{figure_IV.16}
\end{figure}
We show in  Fig.\,\ref{figure_IV.16} 
a contour plot of the $^1D_2$ and $^3F_3$ channels 
for $T_{Lab}$ from 300 MeV to 1 GeV
and off-shell momenta $q=k_{off}$ from 0 to 7 fm$^{-1}$. The  Nijmegen-II 
and inversion reference potentials are  used with  these calculations. 

The $^1D_2$ and $^3F_3$ channels were selected specifically as 
they are noticeably influenced by the $\Delta$ (1232) resonance. 
They also have the
most drastic variations of optical potentials with the 
choice of reference potential.
The results support our  expectations,  associated with strong 
energy dependences and/or large differences 
of experimental and reference potential phase shifts,  which led 
to a scattering
scheme shown in Fig.\,\ref{figure_II.9}. It is not difficult to foresee
great problems in  microscopic analyses which attempt  to describe the 
interferences between back-ground and resonance scattering, and 
which aim for a unique  high quality result. 

For energies above 1 GeV no obvious resonance effect can be identified with
elastic scattering phase shifts. However, this smoothness does not imply that
the off-shell t-matrices are independent of the  choice of optical potential
parameterization. 
\begin{figure}[tbp]
\centering
\epsfig{file=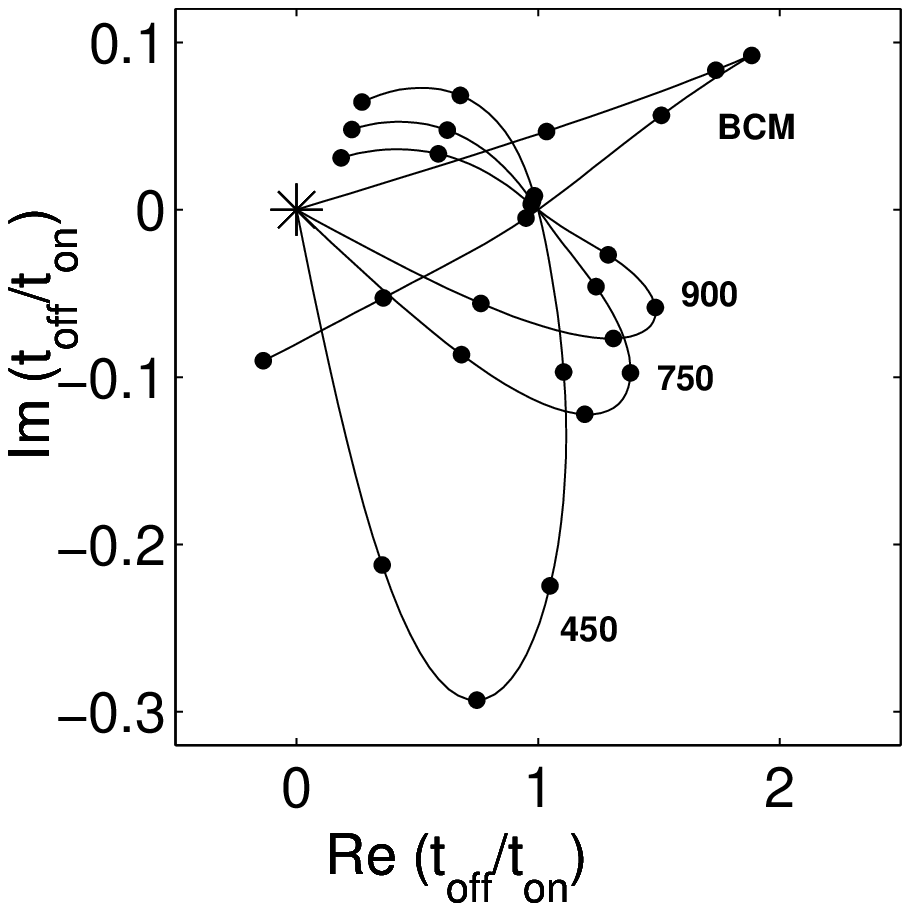,scale=1.0}
\caption{Kowalski-Noyes f-ratios for the $^3P_0$ channel 
calculated with Nijmegen-II as reference potential and 
optical potentials using HO $\hbar\omega=450,\ 750,\ 900$ MeV, 
and a normalized edge function (BCM) separable 
form factor. The dots on the curves fall onto integer $k_{off}$ values 
and  $k_{on}$ is that for $T_{Lab}=2$ GeV.}
\label{figure_IV.17}
\end{figure} 
In Fig.\,\ref{figure_IV.17} we show in the complex plane 
several  Kowalski--Noyes f-ratios for the $^3P_0$ channel.
In three cases we used HO form factors with  $\hbar\omega=450$, 
750 and 900 MeV, 
and in one case we used  a {\em normalized edge function} 
($r_0=0.45$ fm, $h=0.015$ fm)
of Sect.\,III.B as boundary condition. Quite similar results 
were found for the other channels
and the  off-shell differences between these results are significant. 
But the influences of such large and obvious off-shell
 differences disappear when those
off-shell t-matrices are used in  few- and many-body
 calculations \cite{Amo00}. 
It is generally argued that
only near on-shell values enter in few- and many-body
 calculations and symmetric  sampling around the
on-shell point  implies that any effects of such differences 
 are annulled. 
Thus we do not expect medium energy few- and many-body
 calculations to be more revealing
than were the results of  calculations at low energy. 
We consider it not opportune  to seek or nominate
a preference for any of the off-shell t-matrices or 
 particular form factors.
 

\section{Summary and  conclusions}
Diverse nucleon-nucleon r-space potentials, that yield quality 
fits to NN scattering 
phase shifts for energies below 300 MeV, have been extended to
 be NN optical potentials
from which the SP00 phase shift functions to 3 GeV are matched.
 Complex
short range separable potentials, addressed as the optical model 
potential and distinguished from the
real reference potentials, bridge the gap between the experimental 
and reference potential phase shifts. By extending boson exchange
 motivated 
NN potential models
to be optical models
we invoke a new reaction scheme. At medium energy, 300 MeV to 1 GeV,
this approach identifies intrinsic  excitation of isolated  nucleons
without their fusion. At higher energies, and in particular 
for energies $T_{Lab}>2$ GeV, the two  nucleons can fuse into a
 compound system, from
which meson production and other reactions eventuate, as can 
condensation  back 
into the elastic channel. This
view is based upon the character of the $^1S_0$ and $^3P_{0,1}$  
partial wave phase shifts.
Notably  it is the  minimum in the real phase shifts of these channels  
which transform into soft core potentials. The reaction 
volume of the fused 
system fits well within a sphere with  radius  1 fm and
the medium and long range boson exchange contributions 
are small corrections at best. While data at even higher 
energies may indicate a 
similar reaction scheme with the higher partial waves, 
it must be borne in mind
that the centripetal barrier screens that scattering so 
reducing markedly 
the probability of fusion. 

In the 300 MeV to 1 GeV regime, the $\Delta$-resonance
dominates $^1D_2,\ ^3F_3$ and $^3PF_2$ partial waves and all
 reference potentials require
large and strongly energy dependent contributions from the 
 optical potential.
Our results complement the view that this resonance must be 
treated explicitly. 
In our case, the $\Delta$ generates a doorway state to pion
 production and should be treated
as such within the NN potential model generalization.
The separable optical potential was chosen to accommodate
 doorway state formation and decay
within a small energy region. 

The OBE reference potentials presently available  either give
results too far from reality to qualify as background phase shifts  or use
the $\Delta$-resonance in a way that prohibits separation 
from the background. 
However, by dint of their construction, inversion algorithms
 will help resolve these issues. 
The approach is such that one may
start with any desired phase shift function as input. Of these 
any real part may be taken as  the reference potential phase shifts, 
whose use as input to 
Gel'fand--Levitan--Marchenko inversion give the reference 
potentials themselves.
Therewith, the inversion algorithm we have developed herein 
can then be used to determine
the remaining parts of the full NN optical potential. This 
algorithm facilitates 
specification not only of complex separable potentials, appropriate for
specific doorway state effects,  but  also of local 
complex potentials which 
encompass smooth energy dependent processes that contribute 
to medium to
high energy NN data. The geometric attributes of the optical
 model, in particular the
inherent soft core nature of potentials, thus have been 
determined solely from data.
Detailed interpretation of these emergent results, of course,  
must eventuate from  QCD inspired models.

\section*{Acknowledgement} 
This research was supported in part by an ARC grant.
\end{document}